\documentclass[pdflatex,bst/sn-nature]{sn-jnl}

\usepackage{graphicx}%
\usepackage{multirow}%
\usepackage{amsmath,amssymb,amsfonts}%
\usepackage{amsthm}%
\usepackage{mathrsfs}%
\usepackage[title]{appendix}%
\usepackage{xcolor}%
\usepackage{textcomp}%
\usepackage{manyfoot}%
\usepackage{booktabs}%
\usepackage{algorithm}%
\usepackage{algorithmicx}%
\usepackage{algpseudocode}%
\usepackage{listings}%
\usepackage{graphicx}%

\usepackage{float}
\usepackage{tikz,graphics,color,float,epsf,subcaption}
\usepackage{caption}

\usepackage{bm}%
\usepackage{natbib}
\usepackage{tabularx} 
\usepackage{multicol}
\usepackage{array}
\usepackage{multirow}
\usepackage{booktabs}
\def\checkmark{\tikz\fill[scale=0.3](0,.35) -- (.25,0) -- (1,.7) -- (.25,.15) -- cycle;}

\theoremstyle{thmstyleone}%
\theoremstyle{thmstyletwo}%

\theoremstyle{thmstylethree}%

\begin{document}

\title{Multi-Agent Reinforcement Learning for Resources Allocation Optimization: A Survey}


\author[1]{\fnm{Mohamad A. Hady}}\email{mohamad.hady@mymail.unisa.edu.au}

\author*[1]{\fnm{Siyi Hu}}\email{Siyi.Hu@unisa.edu.au}

\author[1]{\fnm{Mahardhika Pratama}}\email{Dhika.Pratama@unisa.edu.au}

\author[1]{\fnm{Jimmy Cao}}\email{jimmy.cao@unisa.edu.au}

\author[1]{\fnm{Ryszard Kowalczyk}}\email{Ryszard.Kowalczyk@unisa.edu.au}

\affil[1]{\orgdiv{STEM}, \orgname{University of South Australia}, \orgaddress{\street{Mawson Lakes Blvd}, \city{Mawson Lakes}, \postcode{5095}, \state{South Australia}, \country{Australia}}}







\abstract{
  Multi-Agent Reinforcement Learning (MARL) has become a powerful framework for numerous real-world applications, modeling distributed decision-making and learning from interactions with complex environments. Resource Allocation Optimization (RAO) benefits significantly from MARL’s ability to tackle dynamic and decentralized contexts. MARL-based approaches are increasingly applied to RAO challenges across sectors playing pivotal roles to Industry 4.0 developments. This survey provides a comprehensive review of recent MARL algorithms for RAO, encompassing core concepts, classifications, and a structured taxonomy. By outlining the current research landscape and identifying primary challenges and future directions, this survey aims to support researchers and practitioners in leveraging MARL’s potential to advance resource allocation solutions.}




\maketitle

\section{Introduction}
Multi-agent reinforcement learning (MARL) has quickly become an essential area of research, providing effective solutions for distributed decision-making in dynamic and decentralized environments. As multiple agents interact and learn in shared settings, MARL addresses the complexities of real-world applications, especially in situations with non-stationary and evolving conditions \cite{ning2024survey,hao2023exploration,nguyen2020deep}. In parallel, Resource Allocation Optimization (RAO) has gained significant attention, as optimizing resource distribution—such as time, energy, network bandwidth, and computational power—can enhance efficiency and effectiveness across a variety of fields \cite{wei2021multi,liu2023energy, allahham2022multi}.

\begin{figure}[t]
\centerline{\includegraphics[width=0.9\textwidth]{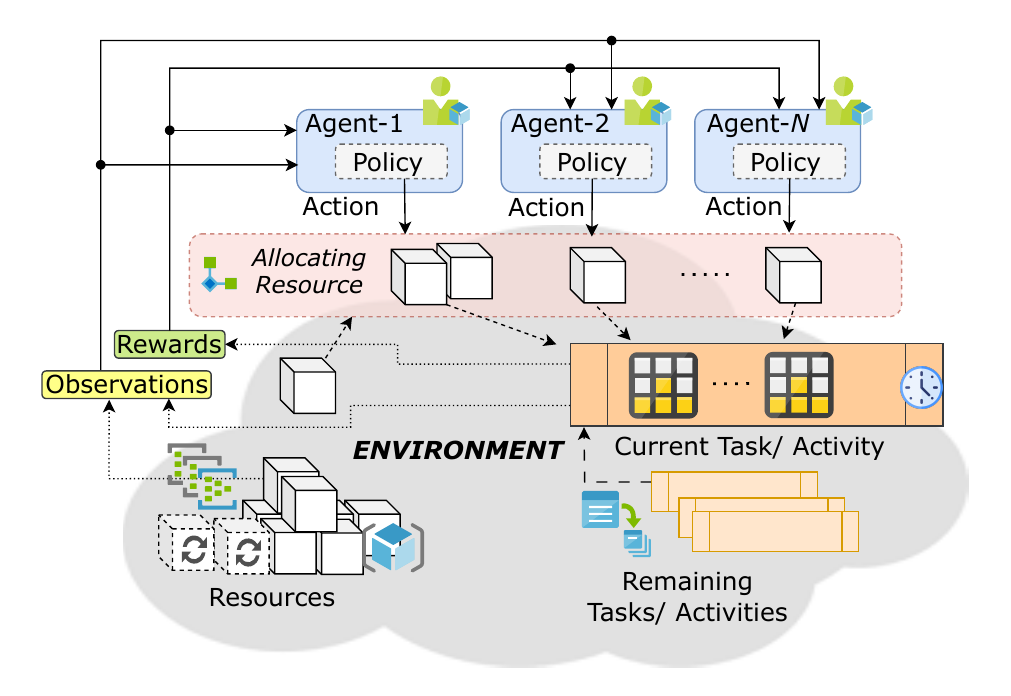}}
\caption{MARL solution for RAO. Any resources can be allocated by several agents to complete tasks or activities. Each agents has its own policy to handle resource allocation determined by the rewards and observations of the whole system states that may come from resources and tasks situations.}
\label{fig:generic_illustration}
\end{figure}

MARL is particularly suited to tackling RAO challenges, as it enables decentralized, adaptive decision-making. This ability is critical in industries like telecommunications, energy management, cloud computing, and transportation, where efficient resource management plays a vital role \cite{wong2023deep,Zabihi2023RLsurvey}. For example, in cloud computing, MARL-based algorithms can optimize resource scheduling and load balancing, leading to improved system performance and reduced costs. Similarly, global supply chains benefit from efficient resource allocation, boosting productivity and reducing expenditures \cite{jiang2009case,ren2022multi}. In power grids, where renewable energy sources are increasingly integrated, MARL enables dynamic energy resource allocation to balance supply and demand \cite{zhang2023multi}. Transportation networks also leverage MARL's adaptive capabilities, with applications such as traffic signal control in cities to alleviate congestion and reduce emissions \cite{wu2020multi}. A generic illustration of MARL solution for RAO framework is provided in Fig. \ref{fig:generic_illustration} that describes how any resources are allocated to a certain task or activity with multi-agent decision makers.

Historically, RAO problems have been addressed through classical optimization and heuristic methods \cite{halabian2019distributed}. However, these methods often lack of the flexibility and scalability required in complex, real-time environments \cite{sarah2023resource}. Reinforcement learning (RL), and particularly MARL, has emerged as a powerful alternative. While RL has proven effective for single-agent optimization problems, MARL extends this to systems with multiple interacting agents, each learning and adapting in real time. This shift from single-agent RL to MARL enables more scalable, decentralized approaches suitable for the increasing complexity of modern RAO scenarios \cite{lei2020deep, noor2020survey,chen2021distributed}.

Despite rising interest in MARL and its RAO applications, there is a lack of comprehensive surveys that focus specifically on this intersection. Existing surveys cover related topics—such as RL in energy systems \cite{yu2021review}, task allocation in multi-robot systems \cite{orr2023multi}, and resource management in wireless networks \cite{feriani2021single}—but do not offer an extensive review of MARL-driven RAO across different industries. This survey fills that gap by providing a focused review on the use of MARL in RAO, with the following primary contributions:

\begin{figure}[t]
\centerline{\includegraphics[width=\textwidth]{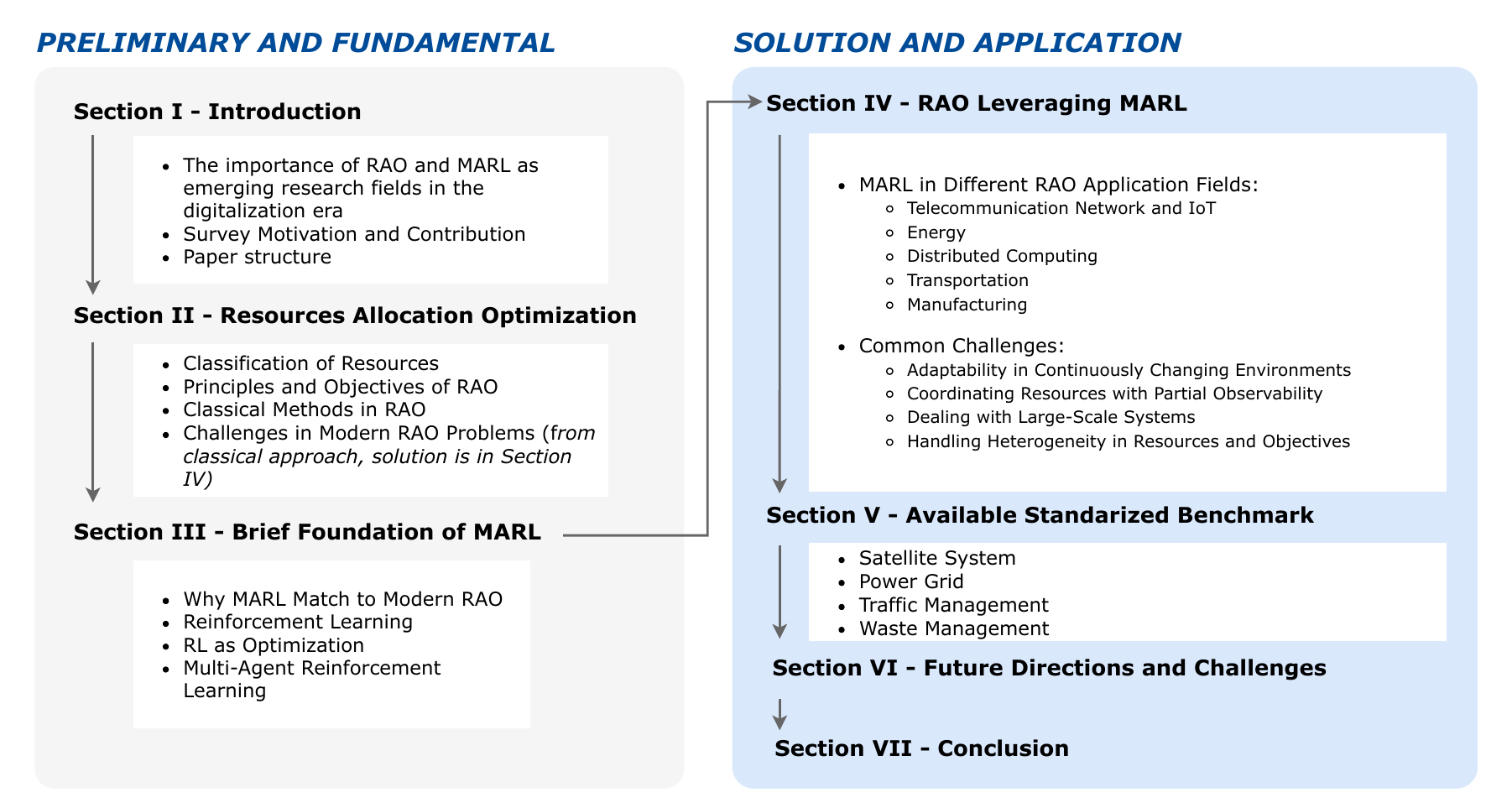}}
\caption{Complete manuscript body structure used in this survey: Preliminary section mainly covers the fundamental of RL and MARL methods. Then, we introduce the concept of RAO, its classical solution and highlight the challenges in the recent trends which can be solved using MARL algorithm and discussed in the RAO leveraging MARL section.}
\label{fig:outline}
\end{figure}

\begin{itemize}
    \item Mapping the current landscape of MARL algorithms and frameworks used in RAO, providing researchers with a consolidated resource.
    \item Offering a systematic review to synthesize advancements, highlight trends, and identify challenges and opportunities unique to MARL applications in RAO.
    \item Categorizing recent literature in MARL training frameworks and their application areas.
    \item Listing real-world benchmarks and available testbeds for RL and MARL algorithm development in RAO.
\end{itemize}

This survey is both timely and necessary, serving as a foundational reference for researchers and practitioners in the field. By focusing on recent developments, selecting high-impact studies, and examining critical aspects such as non-stationary, scalability, agent communication, and coordination, we aim to provide a comprehensive overview of current MARL research in the context of RAO.
The remainder of this survey is organized as follows (see Fig.~\ref{fig:outline}):
\textit{Section 2} categorizes types of resources and reviews classical RAO approaches, including linear programming, heuristic optimization, and game theory. It highlights their applications and limitations in dynamic, large-scale, and decentralized systems.
\textit{Section 3} introduces the fundamentals of reinforcement learning (RL) and extends them to multi-agent reinforcement learning (MARL), with a focus on their applicability to decentralized and dynamic RAO scenarios.
\textit{Section 4} surveys MARL solutions for RAO-related applications and challenges. It reviews the use of MARL in domains such as telecommunications, energy systems, distributed computing, transportation, and manufacturing, and discusses challenges including dynamic environments, partial observability, scalability, and resource heterogeneity. This section also examines key frameworks such as CTCE, DTDE, and CTDE, along with emerging approaches like graph-based MARL.
\textit{Section 5} introduces representative benchmarks used to evaluate MARL performance in RAO settings, including satellite missions, power grids, container management, and traffic systems.
\textit{Section 6} outlines future research directions, emphasizing the need for improved scalability, real-time adaptability, and agent coordination. Potential solutions include hierarchical MARL and mean-field approximation techniques.

\section{Resource Allocation Optimization}

Resource Allocation Optimization (RAO) is a significant area of research across many fields, focusing on the challenge of distributing resources among agents or tasks to improve system efficiency, productivity, or fairness. Resource allocation is the structured distribution of finite resources among tasks or activities to meet a defined objective, such as efficiency or cost. This process is fundamental across various fields, including energy management, cloud computing, manufacturing, and telecommunications, etc. The effective allocation requires identification of available resources, assessment of task demands, and allocation of resources in a way that aligns with system goals, often under constraints such as time, budget, or capacity \cite{feriani2021single}. The primary objective of RAO is to allocate resources as effectively as possible, balancing competing demands to optimize specific goals, such as minimizing delays, maximizing throughput, reducing costs, or ensuring fairness among users \cite{tang2015resource}. In industries like telecommunications, cloud computing, energy distribution, manufacturing, and transportation, efficient allocation of resources such as bandwidth, computational power, energy, and physical assets is crucial to maintaining high performance while minimizing costs and reducing waste. As systems become more complex and interconnected with advancements like the Internet of Things (IoT) and digitalization in large-scale distributed systems, the demand for effective resource allocation becomes even more critical.

\subsection{Principles and Objectives of RAO}

Resource location introduces additional complexity in allocation strategies. Non-distributed resources are concentrated in a single location or managed under a centralized authority. While these systems minimize communication overhead and latency, they often face scalability challenges and are susceptible to single points of failure. Conversely, distributed resources are spread across multiple locations or nodes and require coordination across interconnected systems. This setup offers scalability, fault tolerance, and flexibility, making distributed systems ideal for dynamic environments such as cloud computing, multi-robot coordination, and smart grids. However, distributed systems also pose challenges like communication overhead and coordination latency, necessitating advanced algorithms for efficient resource management \cite{zhang2023resource, huang2023scheduling}. 

\subsubsection{Resource Allocation Process}


In practice, resource allocation strategies are guided by specific characteristics of the resources themselves. For instance, in distributed cloud computing, resources like processing power are spread across multiple servers, necessitating allocation methods that balance load across the network to enhance response times and availability \cite{jiang2015survey}. In contrast, centralized resources in manufacturing, such as machinery, are managed within a single facility to minimize downtime and improve throughput. For divisible resources, such as bandwidth in telecommunications, allocations can vary to meet user demand, reducing latency and enhancing throughput. Indivisible resources, such as individual machines or personnel, require discrete allocation, where entire units are assigned based on task requirements. Finally, renewable resources, like energy from solar panels, are allocated cyclically to maintain availability without depletion, while non-renewable resources, such as a fixed budget, require careful allocation to support long-term goals \cite{NEURIPS2021_1a672771}.

\paragraph{Problem Formulation:} 

Consider a set of tasks indexed by \( i \), where \( i = 1, 2, \dots, n \). Let \( x_i \) denote the amount of resource allocated to the \( i \)-th task. 
The total amount of resource available, \( N \), constrains the allocation as:
\(
\sum_{i=1}^{n} x_i \leq N.
\)
Additionally, task-specific limits on resource allocation, represented by lower and upper bounds \( l_i \) and \( u_i \), are applied as:
\(
l_i \leq x_i \leq u_i, i = 1, 2, \dots, n.
\)
These constraints can be incorporated directly into the optimization model, allowing for efficient adjustments according to operational needs.

\subsubsection{Objective of RAO}

The objective function for RAO, denoted \( f(x_1, x_2, \dots, x_n) \), is formulated to achieve optimal resource distribution by minimizing costs or maximizing benefits. This can be represented as follows \cite{ibaraki1988resource,ushakov2013optimal}:

\begin{equation}
\begin{aligned}
    \text{maximizing} \; f(x_1&, x_2, \dots, x_n) \\
    \text{subject to} \quad \sum_{i=1}^{n} x_i \leq N, &\quad l_i \leq x_i \leq u_i, \; i = 1, 2, \dots, n,
\end{aligned}
\label{genericRAO}
\end{equation}
where \( N \) represents the total available amount of the resource. In cases where the objective is to maximize profit, the problem can be re-framed by minimizing \( -f \), as maximizing \( f \) is equivalent to minimizing its negative.

The structure of the objective function \( f(x_1, x_2, \dots, x_n) \) is often customized to fit the specific requirements of the application, and it may take various forms \cite{patriksson2008survey}:
\begin{itemize}
\item Separable function: A common structure where the objective is expressed as the sum of individual cost functions for each resource, such as \( \sum_{i=1}^{n} f_i(x_i) \).

\item Convex function: When each \( f_i \) is convex, enabling the use of convex optimization techniques for efficiency.

\item Minimax or maximin objectives: In some scenarios, the goal is to minimize the maximum cost, \( \max_i f_i(x_i) \), or maximize the minimum cost, \( \min_i f_i(x_i) \), which can help balance resource allocation across tasks.

\item Fairness-oriented function: A fairness-focused allocation can be achieved by minimizing a function
\( g(\allowbreak \max_i f_i(x_i), \\ \allowbreak \min_i f_i(x_i)) \),
where \( g \) is a non-decreasing function, thereby balancing extremes in resource distribution.
\end{itemize}

\subsection{Resources Properties}

\begin{figure}[t]
    \centerline{\includegraphics[width=\textwidth]{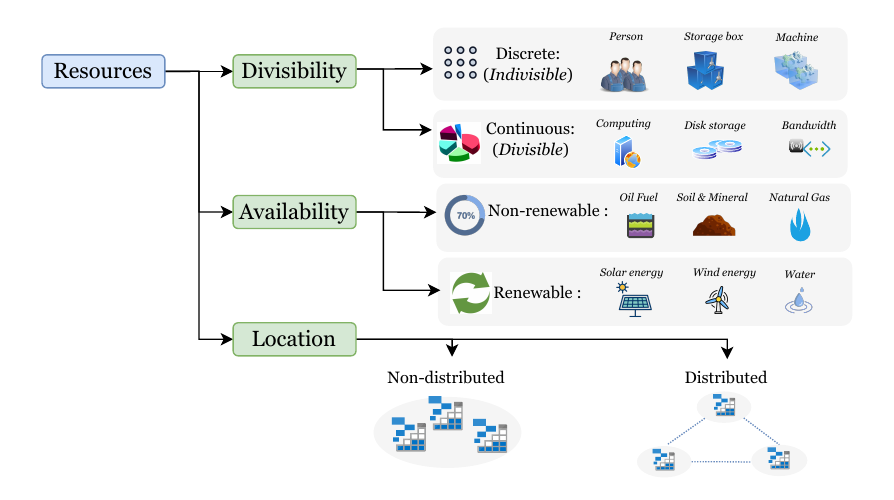}}
    \caption{Resource classification by divisibility, availability, and location properties.}
    \vspace{-10pt}
    \label{fig:resourceType}
\end{figure}

In RAO, the properties of resources play a pivotal role in determining effective allocation strategies. Resources can be broadly classified by their \textit{divisibility}, \textit{duration}, and \textit{location}, each influencing management approaches and system design. These classifications are illustrated in Fig.~\ref{fig:resourceType}. Discrete resources—such as physical machines, servers, or personnel—are indivisible and allocated as whole units. These are typically modeled using Integer Programming (IP), where the resource variable $x_j$ must take non-negative integer values ($x_j \in \mathbb{Z}{\geq 0}$). In contrast, continuous resources—such as bandwidth, processing power, or memory—can be allocated in fractional amounts. These are modeled with real-valued variables ($x_j \in \mathbb{R}{\geq 0}$), enabling more flexible and fine-grained allocation strategies.

The duration of resource availability significantly impacts allocation methods, as resources may be either renewable or non-renewable. Renewable resources—such as solar energy or CPU time—replenish over time and require time-dependent constraints for effective management ($x_i(t) \leq R_i(t),\ \forall t$). In contrast, non-renewable resources—such as budgets, storage, or fossil fuels—are finite and subject to cumulative upper-bound constraints ($\sum_{t} x_i(t) \leq R_i$). These distinctions are essential for modeling real-world systems, where renewable and non-renewable resources often coexist and must be managed in a coordinated and efficient manner. Recognizing the temporal nature of resource availability is therefore critical for designing appropriate allocation strategies.

Effectively managing resource allocation requires a structured approach to align resources with the specific demands of a system. This process involves understanding the type of resources, whether distributed or non-distributed and the needs of the agents or processes involved. Distributed resources, which are common in cloud computing, allow for flexible and scalable allocation across users and systems, though they introduce challenges in managing latency and coordination \cite{Halabian2019, sadatdiynov2023review}. In contrast, non-distributed resources, typical in manufacturing, often rely on a centralized allocation framework that provides direct oversight but may experience bottlenecks as demands increase.

Given these distinctions, the allocation strategy should be adapted to the resource type to maintain system efficiency. This approach requires assessing availability, demand, and constraints to ensure that resources are distributed effectively to meet system objectives, such as maximizing throughput or minimizing idle time. The following section outlines a general process for resource allocation, demonstrating how these principles apply to real-world scenarios. Fig. \ref{fig:resourceType} illustrates the distribution of resources in different properties.

\subsection{Classical Methods in RAO}

This section describes several classical methods to understand their limitations and potential applications in RAO (see the full list of methods in Table \ref{table:classical}).
 \begin{table}[]
 \caption{Summary of classical approaches for RAO Solution}
 \label{table:classical}
\begin{tabular}{|p{3.5cm}|p{3cm}|p{5cm}|}
\hline
\textbf{Category}           & \textbf{Algorithm}                & \textbf{Representative Work}                           \\ \hline
\multirow{2}{*}{Linear   Programming}                    & Mixed Integer Linear \newline Programming & \cite{saaty2003allocation}                        \\ \hline
\multirow{12}{*}{Heuristic \ Optimization} & Simulated Annealing               & \cite{spinellis2000large,suman2006survey,attiya2006task,bi2020energy,kosanoglu2024deep}            \\ \cmidrule{2-3} 
                                        & Genetic Algorithms                & \cite{alcaraz2001robust,cardon2000genetic,tseng2017dynamic,gao2020hierarchical,shao2024gails}                                                  \\ \cmidrule{2-3} 
                                        & Particle Swarm \newline Optimization       & \cite{bratton2007defining,gong2012efficient,lin2018hybrid,liu2022cooperative}                                                  \\ \cmidrule{2-3} 
                                        & Fuzzy Logic Based                 & \cite{xu2008autonomic,wu2018dynamic,khan2019hybrid,zhang2021fuzzy}                                                 \\ \hline
\multirow{4}{*}{Game Theory}            & Cooperative Game                  & \cite{khan2006non,zhang2015resource}                                                  \\ \cmidrule{2-3} 
                                        & Non-Cooperative \newline Game                  & \cite{khan2006non,ye2013non}                                                  \\ \hline
\end{tabular}
\end{table}
 \subsubsection{Linear Programming}

Linear Programming (LP) is one of the foundational techniques in optimization, particularly effective for RAO problems with linear relationships between variables. LP optimizes a linear objective function under linear equality and inequality constraints. This makes it useful for traditional RAO problems such as cost minimization, profit maximization, and efficient allocation of resources like time, money, or physical assets \cite{saaty2003allocation}.

While LP is suitable for problems with manageable numbers of variables and constraints, modern systems like cloud computing or telecommunications often involve high-dimensional settings with large numbers of constraints and variables, making LP computationally expensive for large-scale RAO in real-time contexts.

LP is inherently a centralized approach, assuming a single entity with full knowledge of resources, constraints, and objectives. However, many RAO scenarios involve decentralized systems, like multi-agent systems in smart grids or IoT networks, where LP's centralized nature can lead to inefficiencies due to communication overhead and the need to gather global information.

 \subsubsection{Heuristic Optimization}


Heuristic algorithms provide a flexible alternative for RAO by focusing on finding acceptable, near-optimal solutions within a reasonable time frame. These methods are particularly useful for complex, large-scale, or time-sensitive RAO scenarios where exact solutions are impractical. While heuristics do not guarantee an optimal solution, they offer a practical trade-off between accuracy and computational efficiency, making them suitable for real-time applications.

\begin{itemize} 
    \item \textit{Simulated Annealing (SA)}: Inspired by the annealing process, SA probabilistically explores the solution space, refining the search to find near-optimal solutions, especially useful in complex RAO scenarios \cite{kirkpatrick1983optimization, spinellis2000large}. 
    \item \textit{Genetic Algorithms (GA)}: GA uses principles of natural selection to evolve solutions, making it suitable for large or complex RAO problems with vast search spaces, such as those found in multi-agent systems \cite{cardon2000genetic} and cloud resource allocation \cite{tseng2017dynamic}. 
    \item \textit{Particle Swarm Optimization (PSO)}: PSO simulates social behaviors, where particles (potential solutions) adjust based on their experiences and those of their neighbors. This approach is particularly well-suited for distributed optimization problems in RAO \cite{kennedy1995particle}. 
    \item \textit{Fuzzy Logic-Based Algorithms}: These algorithms handle uncertainty in RAO by using linguistic variables and fuzzy rules, providing flexible allocation even when resource demand and availability are imprecise \cite{xu2008autonomic}. 
\end{itemize}

 \subsubsection{Game Theory}

Game theory models RAO as a strategic interaction among agents, each with potentially competing resource needs. In non-cooperative RAO scenarios, each agent aims to maximize its utility independently \cite{khan2006non,ye2013non}. For example, in wireless communication, users share a common spectrum, each aiming to maximize data rates while accounting for interference from others \cite{cesana2008modelling}. The resulting Nash equilibrium represents an optimized allocation where no user can unilaterally improve their outcome, though it may not achieve global optimization.

In cooperative RAO scenarios, agents may form coalitions to share resources in ways that improve system-wide outcomes. Cooperative game theory models these coalitions, with the Shapley value providing a way to distribute gains fairly based on each agent’s contribution to the coalition \cite{khan2006non,zhang2015resource}. This approach is particularly relevant in decentralized RAO, such as multi-agent systems where autonomous agents either compete or cooperate for shared resources \cite{zhang2012resource}. By capturing interactions among agents, game theory provides a structured framework for optimizing resource sharing in settings where individual and group goals intersect.


\subsection{Limitation and Challenge in Classical RAO Approach}
Classical RAO approaches, including Linear Programming, heuristic optimization, and game theory, provide essential tools for resource allocation. However, modern RAO problems present challenges that these methods cannot fully address. As systems become increasingly complex, interconnected, and dynamic, classical methods struggle with issues like scalability, adaptability, and decentralization. Key challenges include:

\begin{itemize}
    \item \textbf{Continuous and Rapid Changing Issues}: Modern RAO environments, such as power grids, cloud computing, and telecommunications networks, are characterized by rapidly changing resource demands and availability. Classical methods, which assume static or semi-static conditions, struggle to adapt to these dynamic settings. For example, cloud computing experiences sudden fluctuations in demand for processing power, storage, and bandwidth due to varying user activities or service request spikes \cite{zhang2023resource}, while power grids must balance supply and demand in real time to mitigate voltage fluctuations through optimal power management \cite{sun2021two, alam2016computational}. Similarly, next-generation mobile communication networks face challenges in real-time adaptability and dynamic load distribution \cite{wang2024load}. Applications such as autonomous vehicles, financial trading, and smart grids require rapid decision-making under strict time constraints, but classical methods, which rely on solving complex equations or iterative processes, often cannot provide near-instant solutions \cite{singh2017dynamic}. Furthermore, many modern systems operate under significant uncertainties, such as unpredictable network congestion in telecommunications or changing energy supplies and demands influenced by external factors \cite{suzuki2022cooperative}.
    
    \item \textbf{Decentralization and Partial Observability}: In many modern RAO settings, such as multi-agent networks, smart grids, and IoT systems, resource allocation decisions are increasingly decentralized and distributed, with agents making independent decisions based on local information \cite{halabian2019distributed, liao2020distributed, hu2020cooperative, hu2024multi}. Classical, centralized approaches are not well-suited for these scenarios, as they assume full system observability and a central entity managing allocation. In decentralized systems, each agent operates with partial knowledge of the environment and must balance its own objectives against those of other agents. Handling partial observability requires advanced, distributed algorithms that can coordinate effectively without relying on global information, reducing the need for extensive communication overhead.
    
    \item \textbf{Scalability Issues}: As systems grow larger, resource allocation complexity escalates, often exceeding the capabilities of classical optimization techniques. Techniques like linear programming and heuristics become computationally infeasible when applied to large-scale systems, such as cloud computing networks, fog computing, or power grids, where thousands or millions of components (e.g., servers, devices, or users) need to share limited resources \cite{costa2022computational, gao2022large, gao2023large}. The computational costs of classical methods scale poorly with the number of variables and constraints, which leads to significant delays and inefficiencies. Many classical approaches rely on centralized control, where a single entity manages resource allocation based on global system information. In large-scale systems, centralized decision-making is impractical due to bandwidth, latency, and processing limitations, causing bottlenecks and long delays. Techniques like mean-field approximations have been developed to address these scalability challenges, though they remain limited in handling highly interconnected systems \cite{yang2018mean, wang2020large}.
    
    \item \textbf{Heterogeneity of Resources and Objectives}: Modern RAO problems often involve diverse resource types and multi-objective optimization, such as minimizing cost while maximizing efficiency and ensuring fairness. Unlike classical problems that assume homogeneous resources, modern systems must consider a range of resource types (e.g., bandwidth, energy, memory) with unique characteristics and constraints \cite{gao2022fast, zhao2023multi}. Additionally, multi-objective optimization requires balancing competing goals, which classical methods often cannot handle without significant adjustments, making them impractical for complex, multi-dimensional RAO tasks.
\end{itemize}

\begin{table}[]
\caption{Classical Approaches vs. MARL in RAO}
\label{tab:comparison-table}

\begin{tabular}{|p{2.5cm}|p{4.5cm}|p{4.5cm}|}
\hline
\multicolumn{1}{|p{2.5cm}|}{\centering \textbf{Aspect}} &
  \multicolumn{1}{p{4cm}|}{\centering \textbf{Limitation of Classical   Approaches}} &
  \multicolumn{1}{p{4cm}|}{\centering \textbf{Advantage of MARL}} \\ \hline
\centering \textbf{Adaptability} &
  Assume   static or predictable systems; not reactive to changes in real-time &
  Agents   adapt through continuous learning and interaction with dynamic environments \\ \hline
\centering \textbf{Scalability} &
  Poor   performance as system size and complexity increase; computationally intensive &
  Distributes   training and execution across agents, enabling better handling of large-scale problems \\ \hline
\centering \textbf{Decentralization} &
  Often   rely on centralized and global system knowledge &
  Supports   decentralized decision-making and coordination among agents \\ \hline
\centering \textbf{Partial Observability} &
  Require   full or simplified system visibility; struggle with uncertainty &
  Designed   to operate effectively with incomplete or local observations using   Dec-POMDP-based frameworks \\ \hline
\centering \textbf{System Heterogeneity} &
  Limited   to homogeneous models; struggle to accommodate diverse agents or tasks &
  Learns   specialized policies per agent, handling heterogeneous roles and capabilities   efficiently \\ \hline
\end{tabular}
\end{table}

The classical approaches often face several limitations to solve resource allocation as summarized in the Table \ref{tab:comparison-table}. Firstly, classical methods typically model and assume static or predictable environments, making them less adaptable to real-time changes, whereas MARL agents continuously learn and adjust to dynamic conditions. Secondly, scalability becomes a major challenge for traditional techniques as the size and complexity of the system grow, while MARL distributes learning across agents, enabling efficient performance even in large-scale settings. Additionally, classical methods generally rely on centralized control and full system visibility, which limits their applicability in decentralized and partially observable environments. MARL, in contrast, supports decentralized decision-making and performs well under local or incomplete information. Finally, conventional models often struggle to handle heterogeneous systems with varying agent roles or capabilities, while MARL can develop specialized policies tailored to diverse agent functions, allowing for more flexible and robust resource allocation.

These limitations arise some challenges underscore the need to move beyond classical approaches toward more advanced, AI-driven RAO methods, with a particular focus on MARL. In the following sections, we first introduce the foundational concepts of RL and the key principles of MARL. We then survey MARL solutions in addressing RAO problems, structured as follows: Section \ref{4.2.1} explores solutions for continuous and rapidly changing issues, Section \ref{4.2.2} examines approaches to decentralization and partial observability, Section \ref{4.2.3} addresses scalability challenges, and Section \ref{4.2.4} reviews strategies for managing heterogeneity in resources and objectives.

\section{MARL Foundations}


In this section, we begin by reviewing the essential concepts of RL, which form the basis for MARL. By understanding RL’s approach to adaptive decision-making and sequential optimization, we can see how these principles extend naturally to multi-agent environments. MARL leverages the decentralized nature of multiple interacting agents to overcome issues of scalability and enable real-time adaptability in resource allocation.

\subsection{Reinforcement Learning as Optimization}


RL is fundamentally suited to address the decision-making needs in RAO by enabling agents to learn optimal resource allocation policies through trial and error in dynamic and uncertain environments. Through interactions with the environment, agents can identify actions that maximize long-term rewards while adapting to immediate changes in resource demands or availability. This interactive learning process aligns well with RAO’s goal of balancing long-term efficiency with real-time demands. The agent’s Q-function estimates the expected cumulative reward for each action in a given state, guiding optimal actions that balance immediate and future outcomes. In RAO, this allows the agent to dynamically allocate resources based on the current system state and its expected impact on future performance \cite{mao2016resource}.


One of RL’s key advantages in RAO is its capacity to adapt to changing conditions \cite{vengerov2007reinforcement}. Real-world RAO applications often involve unpredictable fluctuations, such as varying workloads in cloud computing or shifts in energy demand in smart grids. RL’s continuous learning framework allows agents to adjust allocation strategies in real time, accommodating shifts in resource availability or demand. Importantly, RL also supports decision-making under uncertainty, allowing agents to explore allocation strategies even without full knowledge of future requirements or the actions of other agents.

Reinforcement Learning (RL) is a machine learning paradigm where an agent learns to make sequential decisions by interacting with an environment to maximize cumulative rewards over time \cite{sutton2018reinforcement}. In RL, the agent’s objective is to learn a policy (a mapping from states to actions) that optimizes long-term rewards, adjusting its actions based on feedback from the environment.



\subsubsection{Markov Decision Process}

The RL framework is formally represented as a Markov Decision Process (MDP), which models the environment, agent actions, state transitions, and rewards \cite{sutton1988learning}. This structure allows agents to dynamically adapt their actions based on observed states and obtained rewards, offering potential for real-time, flexible resource allocation in static or semi-static conditions.
An MDP is defined by a tuple \((S, A, p, r, \gamma)\). Here, \( S \) represents the set of possible states, while \( A \) denotes the set of actions available to the agent. The transition dynamics are modeled by the probability function \( p(s'|s, a) \), which describes the probability of moving from state \( s \) to state \( s' \) after taking action \( a \). The immediate reward, \( r(s, a, s') \), captures the reward received from transitioning from \( s \) to \( s' \) by action \( a \). The discount factor \( \gamma \in [0, 1] \) determines the relative importance of future rewards in the optimization process, weighting immediate rewards more heavily when closer to zero.

\subsubsection{Learning Objective and Bellman Equation}
In an MDP, the objective is to find an optimal policy \( \pi^* \), that maximizes the expected cumulative reward, or return, over time. The return at time step \( t \), denoted \( R_t \), is expressed as the sum of discounted rewards \( R_t = \sum_{i=0}^{\infty} \gamma^i r_{t+i+1}.
\)
To guide its decisions, the agent relies on value functions, specifically the state value function \( V_\pi(s) \) and the state-action value function \( Q_\pi(s, a) \). The state value function, \( V_\pi(s) = \mathbb{E}_\pi \left[ R_t | S_t = s \right] \), represents the expected return when beginning in state \( s \) and following policy \( \pi \).
The state-action value function, \( Q_\pi(s, a) = \mathbb{E}_\pi \left[ R_t | S_t = s, A_t = a \right] \), provides the expected return when starting from state \( s \), taking action \( a \), and then following policy \( \pi \).
These value functions are recursively defined by the Bellman equations, which relate the current value of a state to the expected value of future states and rewards. The Bellman equation for the state value function \( V_\pi(s) \) is given by:
\begin{equation}
V_\pi(s) = \sum_{a} \pi(a|s) \sum_{s'} p(s'|s, a) \left[ r(s, a, s') + \gamma V_\pi(s') \right].
\label{vbell}
\end{equation}
The state value function \( V_\pi(s) \) can alternatively be expressed in terms of the action-value function \( Q_\pi(s, a) \) as:
\begin{equation}
Q_\pi(s, a) = \sum_{s'} p(s'|s, a) \left[ r(s, a, s') + \gamma \sum_{a'} \pi(a'|s') Q_\pi(s', a') \right].
\label{qbell}
\end{equation}
The policy \( \pi \) aims to maximize this equation, yielding the optimal state-action value \( Q^*(s, a) \) across all states and actions.




\subsubsection{Deep RL}

As environments become increasingly complex and involve more factors, traditional methods like SARSA and Q-learning struggle with scalability and stability. This has led to the development of advanced algorithms such as Deep Q Networks (DQN), which approximate the Q-value function with a neural network \( Q(s, a; \theta) \), where \( \theta \) represents the network parameters. To stabilize training, DQN introduces Experience Replay and a target Q-network to decouple updates. The target Q-value and the corresponding loss function are defined together as:
\begin{equation}
    L(\theta) = \mathbb{E}\left[\left(r + \gamma \max_{a'} Q(s', a'; \theta') - Q(s, a; \theta)\right)^2\right].
\end{equation}
Enhancements like Double DQN \cite{van2016deep} and Dueling DQN \cite{wang2016dueling} improve performance by reducing overestimation bias and separating state values from action advantages.

Policy Gradient (PG) methods optimize the policy by directly maximizing the expected return \( J(\theta) \) using Monte Carlo estimates. In the vanilla PG approach, the gradient is given by:
\begin{equation}
    \nabla_\theta J(\theta) = \mathbb{E}_\pi\left[\sum_{t=0}^{T} \nabla_\theta \log \pi(a_t | s_t; \theta) G_t\right],
\end{equation}
where \( G_t \) is the cumulative reward from time step \( t \) onward. 
Actor-Critic methods \cite{konda1999actor} combine Policy Gradient (PG) methods with value function estimation, where the actor updates the policy parameters, and the critic evaluates the value function.  Advantage Actor-Critic (A2C) \cite{mnih2016asynchronous} further reduces variance by using the advantage function, \( A(s, a) = Q(s, a) - V(s) \), instead of the value function.

To maintain stability, Trust Region Policy Optimization (TRPO) \cite{schulman2015trust} and Proximal Policy Optimization (PPO) \cite{schulman2017PPO} constrain updates to prevent large deviations from the current policy. TRPO achieves this by limiting KL divergence while PPO uses a clipped objective.



\subsection{Multi-Agent Reinforcement Learning}

MARL extends the reinforcement learning framework to environments where multiple agents interact and learn concurrently within a shared space \cite{bu2008comprehensive}. While standard RL methods focus on optimizing the cumulative reward of a single agent, MARL addresses the complexities arising from multiple autonomous agents learning and adapting simultaneously. This is especially relevant for RAO in large-scale, decentralized, and dynamic systems where agents must cooperate or compete to manage limited resources effectively.

\subsubsection{Why MARL Match to Modern RAO}

The capabilities of MARL make it highly effective for addressing the core challenges in modern RAO. Its decentralized training and execution paradigms enable agents to learn and act autonomously, reducing the computational bottlenecks and data flow issues inherent in centralized systems \cite{ma2024efficient}. This scalability is critical for RAO applications that span large-scale networks, such as telecommunications or smart grids, where centralized solutions fall short.

Additionally, MARL’s continuous learning processes allow agents to update their policies based on real-time feedback, making the approach highly adaptable to rapidly changing environments, such as fluctuating energy demands in power grids or dynamic workloads in cloud computing. In RAO settings that involve multi-agent systems, such as multi-robot teams or IoT networks, MARL also supports decentralized operations \cite{zhou2023distributed}. This is especially useful under the CTDE paradigm, where agents rely on local observations and make decentralized decisions, addressing the need for distributed control in large-scale systems with limited information \cite{charbonnier2022scalable}.

MARL is also well-suited to handle heterogeneous resources and diverse objectives. It allows for multi-objective optimization, enabling agents to balance local goals with overarching system objectives—a crucial feature for systems requiring distinct allocation strategies for varied resources like bandwidth, energy, and memory \cite{xiao2023multi}. Moreover, MARL’s support for stochastic games and Dec-POMDPs equips agents to make informed decisions in uncertain environments\cite{nguyen2020deep}, a necessity in RAO tasks with unpredictable variables such as fluctuating network loads in telecommunications or variable energy supplies in smart grids.

By providing a structured and adaptable framework, MARL offers a decentralized, scalable, and responsive approach to the demands of dynamic and complex resource allocation environments \cite{ning2024survey}. The following sections will delve into specific MARL algorithms and methods tailored to these challenges, examining their potential to meet the unique requirements of modern RAO applications.

\subsubsection{Stochastic Games}

In multi-agent systems (MAS), agents operate within cooperative, competitive, or mixed environments. These interactions are captured through the formalism of \textit{Stochastic Games}, a generalization of Markov Decision Processes (MDPs) that accommodates multi-agent dynamics \cite{hu2003nash}. A stochastic game, or multi-agent MDP (MA-MDP), involving $N$ agents is defined by the tuple:
\(\langle S, \{A_i\}_{i=1}^{N}, T, \{r_i\}_{i=1}^{N} \rangle,\)
where $S$ represents the set of states, $A_i$ is the set of actions available to agent $i$, forming the joint action set $A = A_1 \times \cdots \times A_N$, $T: S \times A \times S \rightarrow [0,1]$ is the state transition function, and $r_i: S \times A \rightarrow \mathbb{R}$ is the reward function for each agent $i$.

In stochastic games, the state transitions depend on the joint action \( \mathbf{a} = (a_1, \dots, a_N) \) of all agents, which is critical for addressing dynamic RAO challenges. In fully cooperative settings, all agents share a single reward function, aligning with the goal of optimizing resource allocation for a common objective. In contrast, competitive settings introduce conflicting objectives (e.g., in zero-sum games where \( r_1 + r_2 = 0 \) for two agents), while mixed settings involve a combination of cooperative and competitive elements, providing flexibility to model various RAO scenarios.

\subsubsection{Partially Observable MAS and Dec-POMDP}

In practical multi-agent systems, agents frequently face \textit{partial observability}, where each agent only has limited information about the environment's state, introducing additional complexity due to uncertainty. To address this, \textit{Partially Observable Markov Decision Processes (POMDPs)} extend the MDP framework, allowing agents to make decisions based on partial and uncertain observations. 

The multi-agent version, known as the \textit{Decentralized Partially Observable Markov Decision Process (Dec-POMDP)}, is well-suited for decentralized RAO scenarios where agents operate based on localized information \cite{oliehoek2016concise}. A Dec-POMDP is defined by the tuple:
\(
\langle S, \{A_i\}_{i=1}^{N}, T, r, \{O_i\}_{i=1}^{N}, O, N, \gamma \rangle
\),
where $S$ represents the set of environment states, and each agent $i$ has an action space $A_i$, forming the joint action space $A = A_1 \times \cdots \times A_N$ for $N$ agents. The state transition function $T: S \times A \times S \rightarrow [0,1]$ describes the probability of transitioning from state $s$ to $s'$ given the joint action $\mathbf{a} = (a_1, \dots, a_N)$. The global reward function $r: S \times A \rightarrow \mathbb{R}$ provides feedback based on the joint actions. Each agent $i$ has an observation space $O_i$, and the joint observation space is $O = O_1 \times \cdots \times O_N$. The observation function $O: S \times A \times O \rightarrow [0,1]$ gives the probability of agent $i$ receiving observation $o_i$ given state $s$ and joint action $\mathbf{a}$. The number of agents is $N$, and $\gamma \in [0,1]$ is the discount factor that determines the importance of future rewards.

Dec-POMDPs inherently support cooperative settings, aligning with RAO tasks that rely on collaboration for optimal resource distribution. Cooperative MARL enables agents to work collectively to achieve system-wide objectives, addressing modern challenges like scalability by distributing decision-making across agents. The decentralization inherent in Dec-POMDPs addresses the impracticality of central control in large-scale systems, while partial observability reflects the realistic limitations on agents’ knowledge in decentralized RAO settings.




\subsubsection{Learning Paradigms}

\begin{figure}[t]
    \centering
    \begin{subfigure}[b]{0.3\textwidth}
        \includegraphics[width=\textwidth]{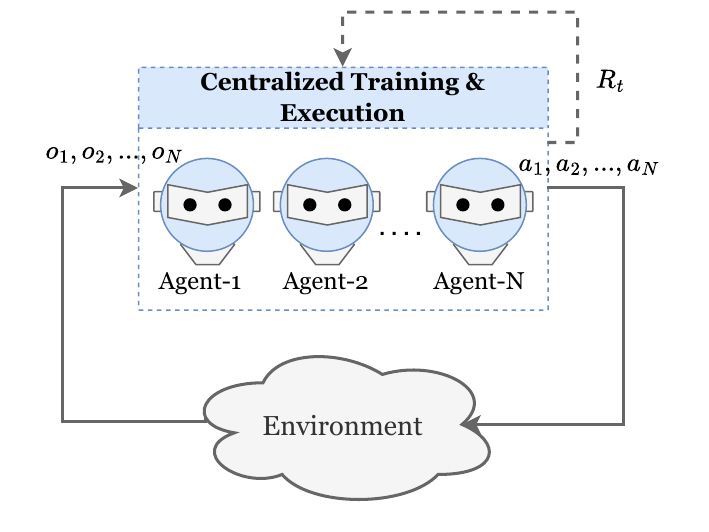}
        \caption{Centralized Training and Centralized Execution (CTCE)}
        \label{fig:ctce}
    \end{subfigure}
    \hfill
    \begin{subfigure}[b]{0.3\textwidth}
        \includegraphics[width=\textwidth]{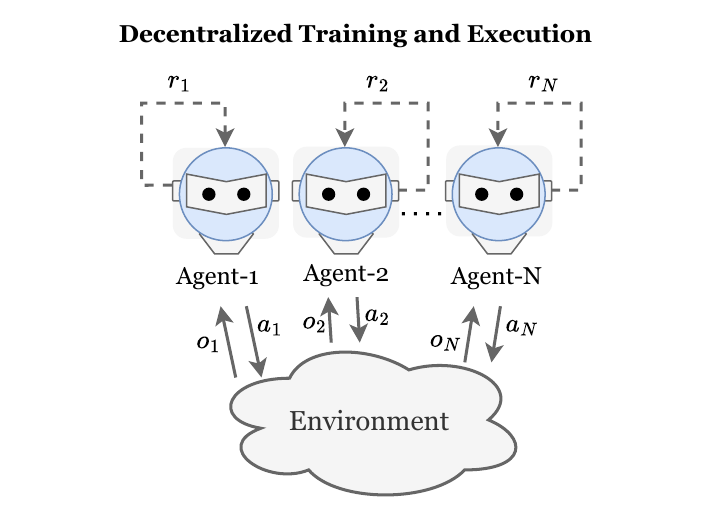}
        \caption{Decentralized Training and Decentralized Execution (DTDE) \citep{wen2021dtde}}
        \label{fig:dtde}
    \end{subfigure}
    \hfill
    \begin{subfigure}[b]{0.3\textwidth}
        \includegraphics[width=\textwidth]{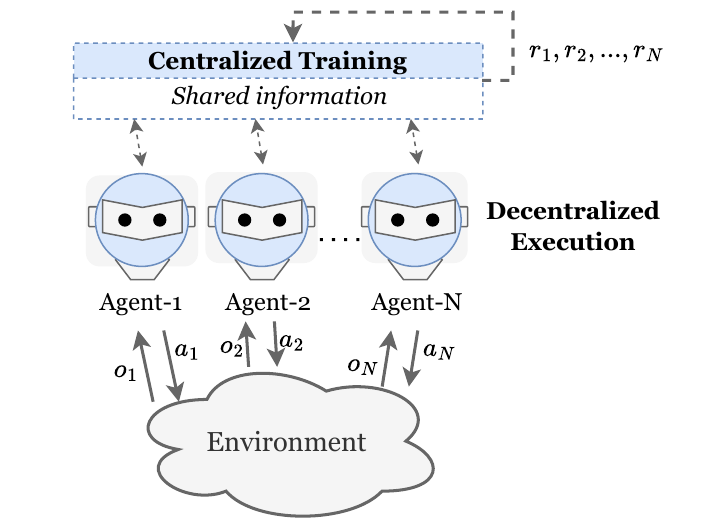}
        \caption{Centralized Training and Decentralized Execution (CTDE)}
        \label{fig:ctde}
    \end{subfigure}
    \caption{Comparison of Training and Execution Paradigms in MARL: (a) CTCE is a fully centralized framework which combines all observations and actions of the agents into a joint observation-action space. (b) DTDE is a term for fully decentralized setting that treats all agents independently with their own observation, action, and reward. (c) CTDE framework has training in a centralized manner with information from other agent, then deploy the trained policy to each agent independently.}
    \label{fig:three_models}
\end{figure}

The training of agents in MARL follows three main paradigms. Each paradigm presents unique benefits and trade-offs for modern RAO, depending on the scale, distribution, and interdependence of resources. Figure \ref{fig:three_models} illustrates the structural differences among these paradigms.

\paragraph{Centralized Training and Centralized Execution (CTCE)}
In CTCE paradigm, a centralized controller determines the actions for all agents based on the global state, optimizing coordinated interactions. The controller utilizes a global policy \( \pi_{\text{global}}(s) \) or a value function \( Q_{\text{global}}(s, \mathbf{a}) \) to select the optimal joint action \( \mathbf{a} \), either by setting \( \mathbf{a} = \pi_{\text{global}}(s) \) or by choosing \( \mathbf{a} = \arg\max_{\mathbf{a}'} Q_{\text{global}}(s, \mathbf{a}') \). While CTCE ensures coordinated actions, scalability can be challenging as the number of agents increases, due to the exponential growth of the joint action space and the communication demands required for real-time access to the global state. CTCE is best suited to controlled environments with relatively few agents, where centralized control remains practical, while it encounters limitations in larger or more decentralized RAO settings, such as distributed energy networks or cloud systems, where scalability and latency are significant considerations.

\paragraph{Decentralized Training and Decentralized Execution (DTDE)}
In DTDE, each agent operates independently, selecting actions based solely on its local observations. The policy \( \pi_i(o_i) \) maps each agent’s local observation \( o_i \) to an action \( a_i \) as \( a_i = \pi_i(o_i) \), or alternatively, using a local value function \( a_i = \arg\max_{a_i'} Q_i(o_i, a_i') \). This decentralized framework offers advantages in scalability and flexibility, enabling agents to operate autonomously. As such, DTDE is suitable for highly distributed settings, such as sensor networks or fully decentralized RAO scenarios. However, DTDE does not inherently provide coordinated actions, which can result in non-stationary environments where agents must adapt to the evolving policies of others. Independent learning (IL) methods, such as Independent Q-Learning (IQL) and Independent Proximal Policy Optimization (IPPO), are effective for large-scale, distributed tasks but may encounter challenges when coordination or shared objectives are essential, as seen in cooperative RAO contexts.

\paragraph{Centralized Training and Decentralized Execution (CTDE)}
CTDE combines the benefits of centralized training and decentralized execution. During training, agents have access to global state information, which allows for a centralized learning process. A global Q function $Q_{\text{global}}(s, \mathbf{a})$ or value function $V_{\text{global}}(s)$ is used to optimize the joint behavior of agents
During execution, however, each agent follows its individual Q function $Q_i(o_i, a_i)$ or learned policy $\pi_i(o_i)$ based on local observations $o_i$,
allowing for decentralized decision-making without access to the global state or knowledge of other agents' actions. 
Key approaches under CTDE include centralized critic and credit assignment methods. The centralized critic approach (e.g., MADDPG \cite{lowe2017multi} and MAPPO \cite{yu2022surprising}) uses a centralized function during training to evaluate joint actions, while credit assignment methods like Value Decomposition Networks (VDN) \cite{VDNSunehag17} and QMIX \cite{rashid2020monotonic} decompose joint rewards to guide individual agents. CTDE’s flexibility makes it ideal for complex RAO applications, such as managing resources in distributed cloud environments, where agents must independently make allocation decisions while aligning with overall system objectives.

 \section{RAO Leveraging MARL}

MARL offers a powerful approach to tackling modern challenges in RAO, particularly where traditional methods fall short in scalability, dynamic adaptation, and decentralized decision-making. MARL extends reinforcement learning (RL) to involve multiple agents interacting within a shared environment, enabling a collaborative, decentralized approach to resource allocation that adapts to complex and changing conditions \cite{bu2008comprehensive}. In this section, a summarization of the available MARL algorithm for different application is discussed, then followed by the primary challenges addressed in that particular field and application.

\begin{table}[]
\centering
\caption{MARL for RAO in different application and its primary challenge: A = Adaptability, PO = Partial Observability, LS = Large Scale, H = Heterogeneity}
\label{tab:application}
\renewcommand{\arraystretch}{1.2}
\begin{tabular}{p{2.5cm}p{3.5cm}p{2cm}cccc}
\hline
\multirow{2}{=}{\textbf{Field}} &
  \multirow{2}{=}{\textbf{Application}} &
  \multirow{2}{=}{\textbf{MARL Algorithm}} &
  \multicolumn{4}{c}{\textbf{Primary Challenge}} \\ \cmidrule{4-7}
  & & & \textbf{A} & \textbf{PO} & \textbf{LS} & \textbf{H} \\ \hline \hline

\multirow{8}{=}{Telecommunication Network and IoT (Section \ref{app:Tele})} &
  Vehicular Network &
  \begin{tabular}[c]{@{}l@{}}MADQRL \\ MADDPG \\ MAD3QN \end{tabular} &
  \checkmark &
  \checkmark &
  \checkmark &
  - \\ \cmidrule{2-7} 
 &
  Mobile Network &
  \begin{tabular}[c]{@{}l@{}}Graph   Based \\      MADDPG \\      MAPPO \end{tabular} &
  \checkmark &
  - &
  - &
  \checkmark \\ \cmidrule{2-7} 
 &
  Wireless   Network &
  \begin{tabular}[c]{@{}l@{}}MADQRL \\      MAPPO \end{tabular} &
  - &
  \checkmark &
  \checkmark &
  \checkmark \\ \cmidrule{2-7} 
 &
  Computer   Network &
  MADQRL &
  \checkmark &
  - &
  \checkmark &
  - \\ \cmidrule{2-7} 
 &
  IoT Network &
  MAPPO  &
  - &
  \checkmark &
  \checkmark &
  - \\ \hline
\multirow{8}{=}{Energy (Section \ref{app:Energy})} &
  Microgrid &
  \begin{tabular}[c]{@{}l@{}}MATRPO \\      MADDPG \\      MATD3 \end{tabular} &
  \checkmark &
  - &
  - &
  \checkmark \\ \cmidrule{2-7} 
 &
  Smart grid &
  MAPPO  &
  \checkmark &
  - &
  - &
   \\ \cmidrule{2-7} 
 &
  Renewable   Energy &
  \begin{tabular}[c]{@{}l@{}}MAD3QN \\      MATD3 \end{tabular} &
  \checkmark &
  - &
  - &
  - \\ \cmidrule{2-7} 
 &
  Power Distribution &
  \begin{tabular}[c]{@{}l@{}}Graph Based \\      MADDPG \\      MATD3 \end{tabular} &
  \checkmark &
  \checkmark &
  \checkmark &
  - \\ \cmidrule{2-7} 
 &
  Home Energy Managemet &
  MADQRL  &
  \checkmark &
  - &
  \checkmark &
  - \\ \hline
\multirow{8}{=}{Distributed Computing (Section \ref{app:computing})} &
  Satellite Edge Computing &
  MAAC  &
  - &
  - &
  \checkmark &
  - \\ \cmidrule{2-7} 
 &
  Mobile Edge   Computing &
  \begin{tabular}[c]{@{}l@{}}MAD3QN \\      MADDPG \\      MAAC \\      MAPPO \end{tabular} &
  \checkmark &
  - &
  \checkmark &
  \checkmark \\ \cmidrule{2-7} 
 &
  Fog Computing &
  MAAC  &
  \checkmark &
  - &
  - &
  - \\ \cmidrule{2-7} 
 &
  Vehicle Edge Computing &
  \begin{tabular}[c]{@{}l@{}}MAD3QN \\      MAAC \\      MADDPG \\      MAPPO \end{tabular} &
  \checkmark &
  \checkmark &
  - &
  \checkmark \\ \cmidrule{2-7} 
 &
  Vehicle Fog Computing &
  MAAC  &
  \checkmark &
  \checkmark &
  - &
  \checkmark \\ \hline
\multirow{3}{=}{Transportation (Section \ref{app:transportation})} &
  Traffic Management & 
  \begin{tabular}[c]{@{}l@{}}MADDPG \\ MAAC \\MADQRL\\ Graph Based \end{tabular} &
  \checkmark  &
  \checkmark &
  \checkmark  &
   - \\ \cmidrule{2-7} 
 &
  Autonomous Vehicles &
  \begin{tabular}[c]{@{}l@{}}MADDPG \\ MATD3 \\ MAAC \\MADQRL \\ Graph Based \end{tabular} &
  \checkmark  &
  \checkmark &
  \checkmark  &
   - \\ \cmidrule{2-7} 
 &
  Vehicle   Allocation &
  MADDPG  &
  - &
  - &
  \checkmark &
  - \\ \hline
\multirow{2}{=}{Manufacturing (Section \ref{app:manufacturing})} &
  Flexible Job shop &
  \begin{tabular}[c]{@{}l@{}} MAPPO \\Graph   Based \end{tabular} &
  \checkmark &
  - &
  - &
  - \\ \cmidrule{2-7} 
 &
  Cognitive Manufacturing &
  Graph Based  &
  \checkmark &
  - &
  - &
  - \\ \hline
\end{tabular}
\end{table}

 \subsection{MARL for RAO in Different Application Fields}
This section explores how MARL serves as a powerful framework for addressing RAO problems across various domains. We begin by discussing how Reinforcement Learning (RL) functions as an optimization technique capable of handling dynamic and uncertain environments. Building on this foundation, we highlight the advantages of MARL in solving complex RAO tasks that involve multiple decision-making agents operating in decentralized or partially observable settings. We then present how MARL has been applied to RAO challenges in diverse real-world fields such as Telecommunication Network and IoT, Energy, Distributed Computing, Transportation, and Manufacturing. Finally, we identify and analyze key challenges commonly encountered in applying MARL to RAO, including adaptability, coordination, scalability, and heterogeneity in resources (see summary in Table \ref{tab:application}).







\subsubsection{Telecommunications and Network Management}
\label{app:Tele}

\paragraph{Vehicular Network}
Resource allocation in vehicular networks is a critical challenge due to the highly dynamic and decentralized nature of vehicular environments. With the growing demand for intelligent transportation systems, efficient allocation of communication, computing, and storage resources is essential to ensure low latency, high reliability, and optimal system performance. The integration of Multi-Agent Reinforcement Learning (MARL) has emerged as a promising solution for enabling distributed decision-making and adaptability in such dynamic settings. MARL allows vehicles and network nodes to learn cooperative strategies in real-time, handle partial observability, and respond effectively to time-varying network conditions. The reviewed papers demonstrate how MARL significantly enhances adaptability in vehicular network environments. Mostly, including \cite{ji2023multi, li2022federated, parvini2023aoi} incorporate adaptive MARL algorithms such as MAD3QN and MADDPG to dynamically respond to changing network conditions and user demands. Studies like \cite{seid2021multi, zhang2020uav, wu2020multi} further apply these models to UAV-enabled IoT, secure communications, and traffic light control, highlighting the algorithms' ability to handle spatio-temporal variability. Distributed and cooperative setups in \cite{hu2020cooperative} and \cite{zhou2023distributed} exemplify the importance of decentralized decision-making in real-time resource management. Although \cite{zhao2023multi} and \cite{yin2021resource} do not explicitly state adaptability, they adopt advanced MARL frameworks capable of operating in heterogeneous and dynamic vehicular networks. Overall, these works underscore the strength of MARL in providing flexible, context-aware solutions for resource allocation, outperforming static or rule-based conventional methods.

\paragraph{Mobile Network}
The reviewed studies in mobile network environments, particularly under the complexity of mobile network management and heterogeneous architectures. \cite{du2022multi} proposed a graph-based MARL (GA-Net MARL) framework for dynamic resource management in 6G in-X subnetworks, emphasizing the capability of graph structures to capture inter-agent relationships and support scalability. \cite{allahham2022multi} addresses the challenges of network selection and resource allocation in multi-RAT (Radio Access Technology) networks using MADDPG, enabling collaborative decision-making in diverse environments. Meanwhile, \cite{Kim2021} leverages MAPPO for end-to-end network slicing, facilitating efficient and adaptive management of resources across network segments. Collectively, these works underline MARL’s ability to handle the heterogeneity, real-time demands, and large-scale coordination challenges present in next-generation mobile networks.

\paragraph{Wireless Network}
The integration of MARL into wireless networks demonstrates promising capabilities for optimizing dynamic and distributed operations. \cite{naderializadeh2021resource} employs MADQRL to manage wireless resources in a decentralized fashion, effectively tackling the coordination challenges among multiple agents in large-scale environments. In a complementary approach, \cite{guo2020joint} proposes a MAPPO-based framework to jointly optimize handover control and power allocation, highlighting MARL’s strength in handling coupled decision-making problems in wireless networks. Together, these studies showcase MARL's potential to outperform classical approaches by enabling adaptive, scalable, and cooperative solutions for complex wireless resource allocation scenarios.

\paragraph{Computer Network}
In computer network application, \cite{you2020toward} leverages a fully distributed MADQRL framework for packet routing, enabling each agent to make routing decisions independently while adapting to network dynamics. This design reduces reliance on centralized control and enhances scalability. Similarly, \cite{suzuki2022cooperative} adopts a cooperative MADQRL approach for dynamic virtual network allocation under fluctuating traffic demands, demonstrating how MARL can effectively respond to variability in network load while maintaining efficient resource use. Both studies underscore MARL’s ability to enable decentralized, adaptive, and resilient solutions for complex computer networking problems.

\paragraph{IoT Network}
In the IoT Network, \cite{xiao2023multi} proposes a MAPPO-based multi-agent deep reinforcement learning framework to address resource allocation challenges in large-scale IoT networks, specifically tailored for ultra-reliable low-latency communication (URLLC) scenarios. Their approach enables decentralized agents to make real-time decisions under strict QoS requirements while coordinating effectively to handle the scale and complexity of controllable IoT systems. This work highlights how MARL, particularly with MAPPO, can meet the stringent demands of emerging IoT applications that require both scalability and reliability in dynamic environments.

\subsubsection{Energy}
\label{app:Energy}
\paragraph{Microgrid}
Recent research has increasingly leveraged MARL for efficient and scalable energy management in microgrid systems. \cite{xu2024collaborative} proposes a hierarchical trust-region MARL framework (MATRPO) to optimize operations across interconnected multi-energy microgrids, focusing on collaboration and trust-aware learning. \cite{abid2024novel} develops a novel multi-objective optimization strategy employing MADDPG to enhance planning decisions for microgrid resource allocation, balancing multiple operational goals. \cite{zhang2023multi} introduced a distributed control architecture using MATD3 for real-time energy management, enabling flexible coordination among microgrids with varying energy types. Complementarily, \cite{jendoubi2023multi} designs a hierarchical MARL model using MADDPG to support layered decision-making, improving learning efficiency and coordination in complex microgrid systems. These works collectively illustrate the diverse applications of MARL techniques in enhancing energy distribution, planning, and real-time control within smart microgrid infrastructures.

\paragraph{Smart Grid}
In smart grid applications to enhance the efficiency and autonomy of energy systems. \cite{kumari2024multi} introduces a decentralized residential energy management system leveraging Deep Q-Network-based MARL (MADQN), focusing on distributed decision-making for home energy optimization. Their approach enables agents to learn optimal consumption strategies while ensuring grid stability. Meanwhile, \cite{roesch2020smart} applies MAPPO in an industrial smart grid context, where multiple agents coordinate to manage energy flows dynamically and adaptively, reflecting the increasing complexity of industrial power systems. These contributions underscore the effectiveness of MARL in addressing decentralized, real-time control challenges in smart grids.

\paragraph{Renewable Energy}
Recent advances in MARL have significantly contributed to optimizing renewable energy integration across diverse systems. \cite{jayanetti2024multi} proposes a MARL framework based on Multi-Agent Actor-Critic (MAAC) for renewable energy-aware workflow scheduling across distributed cloud data centers, aiming to improve energy efficiency and computational performance. \cite{shen2022multi} develops a MAD3QN-based optimization framework for building energy systems, incorporating renewable sources to balance comfort and energy costs. Chen et al. \cite{chen2022physics} introduces a physics-shielded MARL method utilizing MATD3 for active voltage control in photovoltaic and battery-integrated grids, enhancing both safety and operational stability. These studies showcase the versatility of MARL in addressing the complexities of renewable energy systems.

\paragraph{Power Distribution}
In power distribution systems, MARL has emerged as a powerful tool for decentralized voltage and reactive power control. \cite{hu2024multi} proposes a graph-based MARL approach using a decentralized training and decentralized execution (DTDE) framework for Volt-VAR control, demonstrating scalability and coordination among distributed agents. Wang et al. \cite{NEURIPS2021_1a672771} applies both MADDPG and MATD3 algorithms to enhance active voltage regulation performance in distribution networks, improving adaptability in dynamic environments. \cite{sun2021two} introduces a two-stage Volt/VAR control method using MADDPG for active distribution networks, combining global planning and local reactive power adjustment. These works illustrate the effectiveness of MARL in enabling autonomous, robust, and cooperative voltage control strategies in modern power grids.

\paragraph{Home Energy Management}
In the application of Home Energy Management, MARL has been widely explored to optimize distributed control and enhance residential energy flexibility. \cite{charbonnier2022scalable} proposes a scalable MARL framework based on MAQRL to manage distributed energy resources while preserving user comfort and ensuring scalability across numerous households. Xu et al. \cite{xu2020multi} introduces a data-driven home energy management method employing MADQRL, which efficiently adapts to dynamic consumption patterns and uncertain energy generation in smart homes. Similarly, \cite{ahrarinouri2020multiagent} utilizes MADQRL to enable collaborative energy management among residential buildings, achieving significant improvements in both cost reduction and peak load management. These studies demonstrate the capability of MARL approaches to deliver intelligent, decentralized, and adaptive energy management solutions in residential environments.

\subsubsection{Distributed Computing}
\label{app:computing}
\paragraph{Satellite Edge Computing}
In the domain of distributed computing, especially satellite mobile edge computing (SMEC), managing large-scale task offloading across multiple dynamic nodes is a complex challenge. \cite{zhang2024collaborative} addresses this by proposing a Multi-Agent Actor-Critic (MAAC) based collaborative optimization framework. The approach is designed to be scalable and adaptable, enabling decentralized satellite nodes to efficiently learn offloading strategies under highly dynamic and resource-constrained conditions. By coordinating multiple agents in a shared environment, the solution effectively tackles the scalability issues typical in SMEC scenarios, making it promising for future large-scale satellite-enabled computing systems.

\paragraph{Mobile Edge Computing}
The MARL has emerged as a powerful approach for addressing complex challenges in Mobile Edge Computing (MEC), particularly in large-scale environments where resource allocation, task offloading, and adaptability to dynamic conditions are critical. The following papers highlight the challenges of large-scale resource allocation and adaptability in MEC. \cite{liu2024computation} focuses on computation rate maximization for SCMA-aided edge computing in IoT networks with the MAD3QN algorithm, addressing scalability and adaptation in dynamic environments. \cite{gao2023large} and \cite{gao2022large} tackle large-scale cooperative task offloading and resource allocation in heterogeneous MEC systems, utilizing the MAAC algorithm to manage scalability and adaptability in diverse conditions. Gao et al. \cite{gao2023ddpg} proposes Com-DDPG for task offloading in MEC systems for the internet of vehicles, emphasizing the ability to adapt to information-communication-enhanced environments. \cite{zhao2022multi} presents MATD3 for task offloading in UAV-assisted MEC, focusing on scalable and adaptable solutions for mobile edge networks. \cite{cao2020multiagent} applies MADDPG to address multichannel access and task offloading challenges in Industry 4.0, emphasizing large-scale adaptability. Lastly, \cite{wu2023multi} uses MAPPO for minimizing completion delay and energy consumption in MEC-based Industrial IoT (IIoT) systems, emphasizing scalability in handling large-scale IIoT environments.

\paragraph{Fog Computing}
In the application of Fog Computing, \cite{jain2023qos} focuses on QoS-aware task offloading in a fog computing environment, utilizing the MAAC algorithm to address resource allocation and task management challenges in distributed, resource-constrained fog networks.

\paragraph{Vehicle Edge Computing}
\cite{kang2023cooperative} presents a MAPPO-based approach for cooperative UAV resource allocation and task offloading in hierarchical aerial computing systems. \cite{ju2023joint} proposes a joint secure offloading and resource allocation strategy for vehicular edge computing networks using MADDQN, addressing both security and resource management challenges. \cite{zhu2020multiagent} utilizes the MAAC algorithm for vehicular computation offloading in IoT environments, emphasizing task allocation in vehicular edge computing systems. \cite{zhang2021adaptive} combines adaptive digital twin technology with MADDPG to optimize resource management in vehicular edge computing and networks. These studies highlight the importance of scalability, adaptability, and security in resource allocation and task offloading for vehicle edge computing systems.

\paragraph{Vehicle Fog Computing}
\cite{wei2023many} explores many-to-many task offloading in vehicular fog computing using the MAAC algorithm, addressing the challenge of efficiently managing resource allocation and task offloading between multiple vehicles and fog nodes. Gao et al. \cite{gao2022fast} proposes a fast adaptive task offloading and resource allocation approach in heterogeneous vehicular fog computing systems, also utilizing the MAAC algorithm to enhance the efficiency and adaptability of resource management in dynamic vehicular environments. Both studies emphasize the scalability and flexibility required for efficient resource allocation in vehicular fog computing.

\subsubsection{Transportation}
\label{app:transportation}
\paragraph{Traffic Management}
The following research works explore the application of MARL in traffic management. \cite{zhang2024marlens} presents MARLens, a visual analytics approach to understanding traffic signal control using MADDPG.\cite{chen2023deep} applies MAAC for highway on-ramp merging in mixed traffic to enhance traffic flow. \cite{zeynivand2022traffic} implements MADQRL for traffic flow control, optimizing vehicular movement across networks. \cite{wang2020stmarl} introduces STMARL, a spatio-temporal MARL approach for cooperative traffic light control using a graph-based method. \cite{wang2020large} address large-scale traffic signal control with MADQRL, optimizing traffic management in urban settings. \cite{wu2020multi} utilizes MADDPG for urban traffic light control in vehicular networks, focusing on multi-agent coordination to improve traffic efficiency. These studies demonstrate the versatility and effectiveness of MARL algorithms in optimizing various aspects of traffic management.

\paragraph{Autonomous Vehicle}
MARL is applied to autonomous vehicles for optimizing decision-making, coordination, and resource allocation in various driving environments. MARL approaches enable autonomous vehicles to interact, cooperate, and make decisions in complex, dynamic settings, improving safety, efficiency, and traffic flow. Antonio and Maria-Dolores \cite{antonio2022multi} applies MATD3 to manage connected autonomous vehicles at intersections, optimizing vehicle coordination and traffic flow. \cite{jiandong2021uav} explores UAV cooperative air combat maneuvers, using MAAC for decision-making in autonomous vehicle control scenarios. \cite{chen2021graph} leverages graph neural networks combined with reinforcement learning for multi-agent cooperative control of connected autonomous vehicles, optimizing vehicle coordination in complex networked environments. These studies demonstrate the potential of MARL for enhancing the coordination and decision-making capabilities of autonomous vehicles in both traffic and cooperative tasks.

\paragraph{Vehicle Allocation and Routing}
MARL has shown promising results in the application of autonomous vehicles, particularly for resource allocation and optimization in complex transportation systems. By enabling vehicles to interact and cooperate with each other, MARL approaches can improve routing, task allocation, and overall operational efficiency in autonomous vehicle systems. \cite{ren2022multi} applies MADDPG to optimize vehicle routing in supply chain management, focusing on the allocation and efficient routing of vehicles. Their approach integrates route recorders to enhance coordination and decision-making, ensuring that vehicles are efficiently assigned tasks and follow optimal routes, thereby improving supply chain logistics and transportation efficiency. This study showcases the potential of MARL in optimizing vehicle allocation and routing, with a focus on practical applications in supply chain management.

\subsubsection{Manufacturing}
\label{app:manufacturing}
In manufacturing systems, MARL can significantly improve RAO task by enabling decentralized decision-making among autonomous agents, enhancing flexibility, efficiency, and adaptability in complex production environments.

\paragraph{Flexible Job Shop}
MARL has emerged as an effective approach to solve RAO problems in complex manufacturing process, especially in flexible job shop scheduling. In such systems, multiple agents must make real-time, adaptive decisions to allocate resources, optimize production workflows, and meet dynamic constraints, which is essential for increasing efficiency and reducing costs. \cite{jing2024multi} utilizes a graph-based MARL approach combined with Graph Convolutional Networks (GCN) to optimize flexible job shop scheduling, enhancing the adaptability of agents in handling complex scheduling tasks. \cite{heik2024adaptive} proposes MAPPO to dynamically allocate manufacturing resources, addressing the need for adaptive decision-making under changing production conditions. \cite{zhang2023deepmag} introduces DeepMAG, which integrates deep reinforcement learning with multi-agent graphs for flexible job shop scheduling, allowing agents to adapt to evolving task requirements and machine statuses. \cite{liu2023integration} combines deep reinforcement learning and a multi-agent system to dynamically schedule re-entrant hybrid flow shops, accounting for worker fatigue and skill levels, thus improving adaptability in labor resource allocation. Finally, \cite{zhang2022dynamic} focuses on dynamic job shop scheduling using MAPPO, which enables multi-agent manufacturing systems to adapt to real-time changes and uncertainties in the production process. These studies highlight the crucial role of adaptability in optimizing flexible job shop scheduling through MARL, especially dynamic manufacturing environments.

\paragraph{Cognitive Manufacturing}
Cognitive manufacturing, which integrates AI and machine learning, has significant potential for improving decision-making, resource optimization, and process efficiency in manufacturing systems. MARL plays a key role in cognitive manufacturing by enabling multiple agents to work together in a decentralized manner to learn and adapt to various manufacturing tasks and environmental changes. \cite{zheng2021towards} addresses this challenge by proposing an industrial knowledge graph-based MARL approach for cognitive manufacturing. Their approach enables agents to dynamically learn and adapt to different manufacturing tasks through the use of knowledge graphs, which helps in representing complex relationships between various system components.

\subsection{MARL as a Solution for Modern RAO Challenges}

The complexity of modern RAO has led to the extension of RL to multi-agent settings, enabling each agent to make localized and distributed resource allocation decisions. MARL offers a range of paradigms that effectively address the specific challenges in RAO. By extending traditional reinforcement learning to handle multiple agents within a shared environment, MARL introduces solutions that leverage centralized, decentralized, and hybrid learning paradigms. This flexibility enables MARL to meet scalability, adaptability, coordination, and privacy needs in RAO, making it a valuable framework for modern, complex systems. In this subsection, we discuss further four primary challenges and how it is solved 

\subsubsection{Adaptability in Continuously Changing Environments}
\label{4.2.1}

In dynamic and uncertain environments, RAO must handle continuous changes in resource demands and availability that may often change unpredictably. An RL algorithm can be extended to multi-agent settings by simply combining its observations and actions from multiple agents. It constructs a single RL model with a larger amounts of inputs and outputs in a CTCE or fully-centralized framework. By using the CTCE framework, it allows the central controller to quickly adapt the overall strategy by recalculating optimal allocations based on current conditions. This agility is especially valuable in RAO contexts with fluctuating demands, enabling the system to maintain efficient allocations even as conditions evolve. 

In \cite{jain2023qos}, a fully centralized MARL algorithms has been evaluated based on three different DRL methods such as DQN, DDPG, and SAC. It focused on handling the unpredictability of tasks and maintain the Quality of Service (QoS) requirements of users by considering a highly dynamic requirements, such as: end-to-end latency, energy consumption, task deadline, and priority. The algorithms are trained offline in a resource rich cloud data center. The SAC algorithm outperforms other baseline techniques in terms of  time, energy consumption, utility, execution rate, and aging. A decentralized framework may struggle to adapt to these shifts due to limited information access, resulting in slower response times and potential under-utilization or overloading of resources.

In resource allocation, multiple agents may compete for limited resources, and optimal allocation requires that each agent’s actions are highly coordinated to efficiently achieve the same goal. Centralized systems have access to all the agents' states, actions, and rewards, enabling the learning algorithm to optimize resource allocation globally rather than locally. This framework leads to more optimal resource allocation across the entire system since decisions are made with a comprehensive view of the environment. With the full access to all the agents' information and the entire system's state, a centralized system can make more well-informed decisions. This can be particularly beneficial in a cooperative resource allocation when global knowledge is necessary to avoid any over-use or under-use of resources. Decentralized approaches have a partially observable settings that make it struggles with this issue and leading to conflicts or suboptimal outcomes. A deep reinforcement learning algorithm, called Compounded-Action Actor-Critic (CA2C), has been evaluated to address the trajectory planning problem for the cellular Internet of UAVs in \cite{hu2020cooperative}. The CA2C algorithm is capable of effectively managing agents with complex actions, which involve both continuous and discrete variables. This work implements Deep reinforcement learning algorithms that is well-suited for determining optimal policies for agents in MDPs model that have high-dimensional state spaces. All states and actions are jointly processed in a single central controller. They evaluate the proposed algorithm by comparing it against four baseline algorithms. The proposed CA2C algorithm was shown to outperform four benchmark algorithms in terms of AoI minimization.

A fully centralized systems struggle with the scalability issue as the number of agents increases. The central controller must process all agents' states, actions, and rewards in a single computation resource leading to a dramatic increase in computational complexity and memory requirements. This can make centralized MARL impractical for large-scale resource allocation problems. From the reliability point of view, a centralized system is vulnerable to failures in the central node controller. If the central node fails, the entire system can be disrupted, which reduces the system’s fault tolerance and resilience. This can be particularly risky in mission-critical resource allocation tasks. Since all agents' information must be shared with the central controller, fully centralized systems pose privacy and security risks. In resource allocation scenarios involving sensitive data (e.g., personal data in healthcare \cite{Guindo2012}, financial data in blockchain-based resource management \cite{Yanez2020}), these concerns may make centralized MARL less viable. Therefore, this settings are not widely used in RAO research field.


\subsubsection{Coordinating Resources with Partial Observability}
\label{4.2.2}
In many RAO application, each agent can only observe a limited view of the overall environment (e.g., local resource availability or neighboring agents’ actions). For instance, in a network setting, an agent may only know its local bandwidth but not the network congestion affecting other agents. This fragmented view makes it challenging for agents to make well-informed decisions about resource allocation, often leading to suboptimal use of resources. Partial observability leads to scenarios where agents might allocate resources based on incomplete or outdated information, which can create conflicts, over-allocations, or wasted resources. For instance, an agent might allocate a resource already claimed by another, or under-utilized resources that could have been shared more effectively. 

RAO settings often demand that agents make decisions with limited local information, particularly in IoT networks, multi-robot systems, and other distributed architectures. To optimize resource allocation, agents must coordinate their actions with limited information. However, partial observability inherently limits their ability to understand how their decisions impact others, especially in large systems where agents’ actions can have ripple effects across the environment. Inadequate coordination can result in inefficient policies where agents inadvertently work against each other. Addressing partial observability in resource allocation requires a combination of these techniques to create a more cohesive MARL framework, enabling agents to coordinate effectively despite limited information. This way, agents can make more informed decisions, improving both local and global resource utilization. For example, in a grid computing system, multiple agents might allocate tasks to the same computing node, overloading it and slowing down task processing, while other nodes remain idle. This lack of coordination wastes computational resources and increases processing times. Here, \textbf{DTDE} is highly applicable, as it allows agents to make autonomous decisions based on local observations, thus aligning well with environments where agents have constrained or partial access to global states. The independence afforded by DTDE is crucial in applications where maintaining privacy and minimizing communication costs are essential, such as decentralized task allocation in manufacturing systems \cite{zheng2021towards}. 

In an interconnected multi-energy microgrid optimization, \cite{zhang2023multi} proposed solution to the decentralization problem in microgrid system. The authors implemented MADRL algorithm with an attention mechanism added to the centralized critic to meet local customized energy demands in a form of decentralized execution. In this work, CTDE framework is used to maintain global optimization performance with the coordination of each agents. By using this framework, agents are trained with access to a centralized view of the environment, allowing them to learn optimal strategies that account for the whole system’s needs. During execution, agents act based on their local observations, but they are guided by policies shaped by this global perspective. 

However, in cases where some level of centralized oversight is feasible, CTCE can provide fully coordinated solutions. CTCE is ideal for small-scale RAO tasks with a limited number of agents requiring precise synchronization, such as coordinated robotic tasks in structured environments \cite{jain2023qos}. In this setting, CTCE enables the system to optimize resource use by leveraging a complete, centralized understanding of all agents’ actions and states. 

\subsubsection{Dealing with Large-Scale Systems}
\label{4.2.3}
In centralized or semi-centralized approaches, scaling up the number of agents increases the computational complexity and memory requirements due to the prodigious amount of state and action information that needs to be processed. The complexity grows exponentially as the number of agents increases, making it impractical for real-time resource allocation. Besides, In a DTDE framework, each agent is trained and operates independently, relying only on local observations without needing centralized information \cite{jiandong2021uav,wu2020multi}. In other words, the agents are trained to maximize their local rewards and optimize their policies independently. This independence feature reduces computational overhead, as each agent handles its learning and decision-making based on local states, which is more scalable than a central agent tracking the states and actions of all others. With more agents, there is a risk that individual decision quality may degrade if agents lack sufficient information or coordination. In large-scale systems, maintaining high decision quality is critical to ensure efficient use of resources across all agents \cite{gao2023large}. Through local interactions and learning policies tailored to specific environments, agents can still achieve near-optimal decisions at scale, even without full knowledge of other agents’ actions. This framework can maintain decision quality by focusing on improving each agent's local policy, which contributes to a more stable system-wide resource allocation.

Scalability is a core challenge in large-scale resource allocation systems, where handling numerous agents and their interactions becomes increasingly complex. This problem emerge as the system size, number of agents, or complexity of the resource environment grows. Specifically, scalability issues arise when the algorithms or strategies used to allocate resources cannot efficiently handle an increase in agents or resource demands. In RAO, each agent must choose actions that contribute to optimal resource allocation. As the number of agents or resources increases, the combined action space grows exponentially, leading to a massive increase in the number of possible allocation combinations. Managing this vast space is computationally challenging, and traditional methods can struggle to identify optimal or even near-optimal solutions within a feasible time frame. This problem is particularly pronounced in real-time RAO scenarios, such as network bandwidth allocation or power distribution, where decisions must be made quickly. The impact that can be occurred is the agents might redundantly allocate resources, over commit shared resources, or even leave some resources underutilized due to lack of information about other agents' decisions without effective. This results in inefficiencies and potential bottlenecks in resource distribution. Leveraging MARL frameworks, either DTDE or CTDE, can help agents to learn efficient policies in large-scale and partially observable environments without a central controller. 

To solve scalability in RAO problem, DTDE scales well with a large number of agents. Each agent learns independently, allowing the system to handle more agents without requiring centralized coordination or training. This is particularly beneficial in resource allocation scenarios where there are many resources and agents in distributed energy systems or communication networks. Since agents train and execute independently, DTDE systems do not require the sharing of sensitive information among agents or a central controller. This feature can be crucial in situations like network resource allocation, where privacy and security concerns are high. DTDE allows agents to adapt to local environmental changes or constraints without waiting for a centralized update or communication. This can lead to more dynamic and responsive resource allocation, as agents can adjust in real-time based on localized resource availability and demand. In DTDE, agents are trained to act independently without a centralized controller, reducing the computational burden and making the system scalable. In \cite{gao2023ddpg}, a task offloading based on MARL has been implemented for Information-Communication-Enhanced Mobile Edge Computing for the Internet of Vehicles. This work evaluates LSTM network and a BRNN, Actor Critic MARL in a fully decentralized setup and compares all of the algorithms performance.

\textbf{Graph based MARL} has been evaluated to solve problem in power distribution using DTDE framework in \cite{hu2024multi}. The authors proposed algorithm divides the power distribution system into several regions, each region treated as an agent. Then an MARL was designed by employing  Hierarchical Graph Recurrent Network (HGRN) structure that combines the advantages of Hierarchical Graph Attention (HGAT) and Gated Recurrent Unit (GRU) enables the communication between heterogeneous agents. Another graph based MARL has been implemented in \cite{zheng2021towards} by using industrial knowledge graph (IKG)-based multi-agent reinforcement learning. It is designed to solve the robot task allocation and completion problem in a manufacturing network.

MARL’s capacity for decentralized real-time adaptation is a key advantage here, particularly in \textbf{CTDE} framework, where agents are trained with global insights, yet operate with decentralized policies in real-time. For instance, in energy management systems, CTDE enables agents to adapt to fluctuating demands while maintaining overall grid stability \cite{guo2020joint}. By updating policies based on local observations and previously learned global strategies, agents can quickly react to dynamic conditions without centralized intervention. Furthermore, CTDE’s use of a centralized critic during training mitigates the non-stationary issue that arises from agents learning simultaneously. This is particularly effective in RAO applications like edge computing, where agents (servers or virtual machines) optimize resource allocation across fluctuating workloads \cite{wu2023multi}. By combining centralized training with decentralized adaptability, CTDE balances the need for coordination with the flexibility to handle rapid environmental changes.

By training with centralized information but executing with decentralized policies, the system can scale to a large number of agents and environments. During training, agents can learn to cooperate and avoid conflicts or inefficient resource usage. Decentralized execution allows agents to quickly adapt to local changes in demand or resource availability without waiting for global updates. Another benefit is that CTDE reduces communication overhead during execution, which is crucial in systems where bandwidth or computational power is constrained.

One type of Credit Assignment is developed using Value Decomposition (VD) method. The common algorithm that has been used in RAO problem is QMix. It has been evaluated to solve a problem for vehicular cloudlet in automotive industry \cite{ahmed2023marl}. A global critic (centralized function) is used to evaluate the actions of all agents collectively. The global reward is computed based on the combined agent’s performance, and the centralized system updates a joint action-value function. Currently, VD  has been rarely used in RAO compared to Centralized Critic since it has more abstract cooperative strategy represented by decomposing the joint action-value function $Q_{tot}$ into individual action-value functions $Q_i$ for each agent. The decomposition ensures that the sum or combination of the individual values approximates the global value function. However, there is no guarantee that it will converge to global optimal point during the training phase.

In cloud environments, CTDE can be applied to allocate computing resources (e.g., CPU, memory) among multiple virtual machines or applications. Agents (representing servers or VMs) learn to allocate resources based on system load and application requirements, optimizing for performance and cost. In \cite{wu2023multi}, a task offloading optimization problem for edge computing based Industrial IoT (EIIoT) infrastructure has been solved using MARL algorithm. An actor-critic (AC) DRL framework is adopted to construct a suitable lightweight offloading decision system and optimize the joint completion delay and energy consumption in the EIIoT. 

CTDE MARL can be used to allocate spectrum resources dynamically between different users or devices in a 5G or IoT network. Agents (representing base stations or users) learn to allocate bandwidth efficiently based on energy consumption \cite{meng2020power,nasir2019multi}, traffic demand and channel quality \cite{Kim2021}.

In energy management systems, CTDE MARL can help to allocate power from renewable sources among consumers or storage devices. Agents (representing homes, factories, or storage units) learn to request power optimally to balance supply and demand, minimize waste, and ensure grid stability. MARL has been evaluated an optimization of power allocation for home energy management using MAPPO algorithm \cite{guo2020joint}. In fact, MAPPO is also built upon centralized critic network and it shows competitive performance to solve the task in energy management.

\subsubsection{Handling Heterogeneity in Resources and Objectives}
\label{4.2.4}
In multi-agent systems, agents must balance diverse resource types, availability, and possibly conflicting objectives. Resources often vary widely in their characteristics, utility, and constraints, leading to heterogeneity in their availability, value, and compatibility with different tasks. For instance, in cloud computing, resources like CPU, memory, storage, and GPU have different capacities, costs, and efficiencies, and not all tasks require or benefit from each resource in the same way. In smart grids, energy sources might vary (e.g., solar, wind, fossil fuel), each with unique costs, emissions, and reliability levels. Matching the right resource to the right demand while maintaining system efficiency becomes complex in such diverse resource pools. Resource heterogeneity can create allocation inefficiencies and under-utilization. Allocating a resource poorly suited to a task can lead to degraded performance, over-utilization of high-value resources, and an imbalance where some resources are over-consumed while others are underutilized. Heterogeneity in resources and objectives adds complexity to the task of matching available resources with the diverse needs of agents.  Poor matching leads to inefficiency, where high-demand or specialized resources are wasted on low-priority or incompatible tasks. This can reduce overall performance, degrade system stability, and prevent high-priority tasks from accessing the necessary resources.

Many RAO problems involve multiple types of resources, each with unique allocation goals. \textbf{CTDE} is well-suited for these heterogeneous environments, as it allows centralized training to capture complex interdependencies between diverse resource types. For example, in cloud computing, agents trained with CTDE learn to balance between CPU, memory, and bandwidth resources \cite{wu2023multi}, optimizing across multiple metrics such as energy consumption, latency, and throughput.

In situations with diverse agent goals or conflicting objectives, \textbf{Credit Assignment} within CTDE is particularly valuable. Credit assignment decomposes the joint action-value function into individual agent rewards, enabling each agent to receive feedback on its contribution to the overall objective. Value decomposition methods, such as QMix, can facilitate nuanced resource allocation in scenarios like vehicular cloudlets, where agents (vehicles) must manage bandwidth and energy resources to optimize communication \cite{ahmed2023marl}.

A centralized critic (by using actor-critic methods) or a centralized policy network is used to optimize the overall objective, accounting for interactions between agents as in \cite{wei2023many}.  During training, a centralized critic has access to the global state of the system (e.g., overall network load, resource availability, and actions of all agents). This allows the agents to learn how their individual actions affect overall system performance. Agents can share experiences and learn from each other, which speeds up training and improves coordination. Centralized critics approaches address the non-stationary problem caused by the changing behaviors of agents. These help to stabilize learning by using a shared critic function to evaluate the actions of multiple agents, reducing the complexity of dealing with multiple evolving policies. Once the training is complete, agents operate independently based only on their local observations. They no longer have access to the full state of the environment or the actions of other agents. This decentralized execution is scalable to large environments with multiple agents, as each agent can act autonomously without relying on global information in real time.


 \section{Available RAO Simulators for Benchmarking}
In MARL, a simulator or environment for benchmarking is developed based on Open-AI gymnasium (gym) library as their main framework to be integrated with any common reinforcement learning algorithms. Here, we listed some publicly available benchmarks for RAO task inspired by the real-world application which are commonly used, based on gym library and can be widely used for evaluating MARL algorithm before it is implemented to the real application.

\subsection{Earth Observation Satellite Mission}
\begin{figure}[t]
\centerline{\includegraphics[height=7cm]{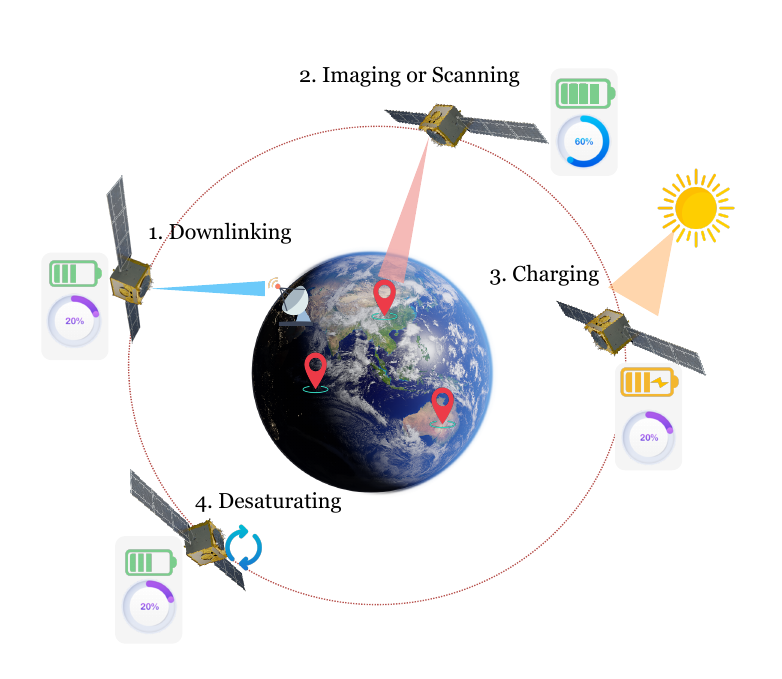}}
\caption{Earth Observation Mission Environment in BSK-RL: The number of satellite can be defined either single satellite or multi-satellite. Each satellite have each actions: 1) Downlinking: Transmit collected data to the ground station with predefined transmission speed and delete the data after successfully downlinked; 2) Imaging or Scanning: Satellite capturing image of a target using optical sensor or scanning any object on the earth surface using radar sensor (with two different payloads and can be used for different tasks); 3) Charging: The satellite is in charging mode and pointing its solar panel towards sun direction to maximize solar energy absorption; 4) Desaturating: There is a condition of the satellite's rotating wheel rotates saturatedly and the speed should be reduced to control their attitude.}
\label{fig:bsk-rl}
\end{figure}

In \cite{stephenson2024bsk}, an open-source Python package for developing and customizing reinforcement learning environment has been designed by the authors to solve spacecraft tasking problems. It integrates Basilisk, a high-performance and high-fidelity spacecraft simulation framework, with the abstractions of satellite tasks and operational goals, all within the standard Gymnasium API wrapper for RL environments (see Fig. \ref{fig:bsk-rl}). This package is specifically designed to support the needs of researchers in reinforcement learning and spacecraft operations. Some works has been done to solve earth observation mission using RL algorithm \cite{herrmann2023reinforcement},\cite{herrmann2024single},\cite{stephenson2024reinforcement} and currently a MARL framework has been provided as an example. This work has an open source codes available here: \url{https://github.com/AVSLab/bsk_rl}.

\subsection{Power Grid Networks}
\begin{figure}[h]
\centerline{\includegraphics[height=5cm]{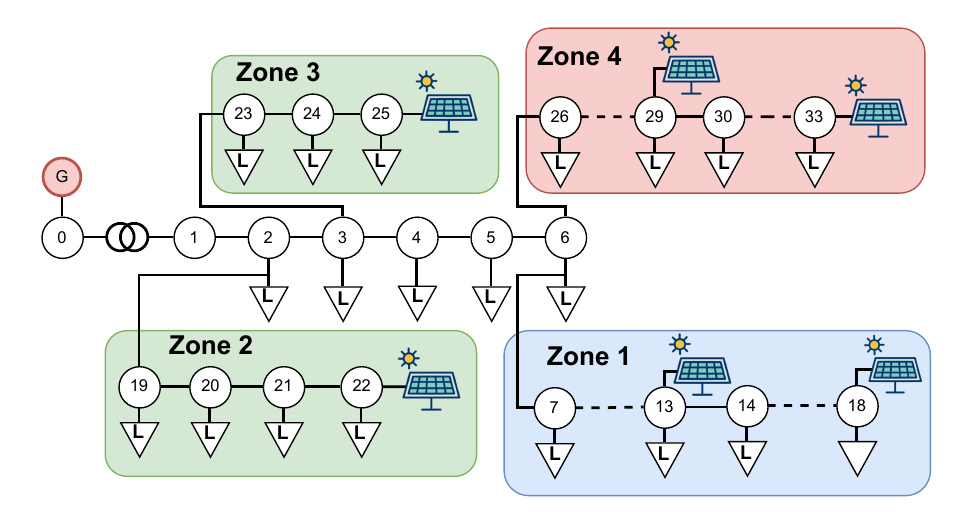}}
\caption{Power Grid Networks environment: This system simulates the energy distribution and control within 4 different zones and 33 Bus Networks. Bus 2-33 voltages should be controlled by the system and Bus 0-1 represent the substation at the main grid with constant voltage and infinite active and reactive power capacity. There are several PV energy sources located in different areas that is challenging to maintain the voltage stability and control. }
\label{fig:PowerGrid}
\end{figure}

A common power management benchmark has been proposed in \cite{NEURIPS2021_1a672771}. This environment introduces a power network problem that provides a compelling yet challenging real-world scenario for the application of MARL as illustrated in Fig \ref{fig:PowerGrid}. The growing trend of de-carbonization is putting significant pressure on power distribution networks. Active voltage control has emerged as a promising solution to alleviate power congestion and enhance voltage stability without requiring additional hardware, by leveraging controllable devices within the network, such as rooftop Photo-voltaic (PV) and Static Var Compensator (SVC). It has been used for a robust evaluation of MARL algorithms in \cite{guo2022towards}, cooperative MARL with individual global max \cite{hong2022rethinking}, dynamic MARL algorithm configuration \cite{xue2022multi}, some power networks research \cite{chen2022physics,lu2023deep}, etc. These devices are numerous and spread across wide geographic areas, making MARL an ideal approach. The codes of this work is publicly available here: \url{https://github.com/Future-Power-Networks/MAPDN}.

\subsection{Traffic Management}
CityFlow is a multi-agent reinforcement learning (MARL) environment developed for large-scale urban traffic management \citep{zhang2019cityflow}. It tackles the complex task of traffic signal control, which requires real-time adaptation to dynamic traffic conditions and coordination among thousands of agents, such as vehicles and pedestrians. CityFlow features fundamentally optimized data structures and efficient algorithms, enabling high-speed, city-wide simulations. It supports flexible road network and traffic flow configurations based on both synthetic and real-world data, and includes a user-friendly interface for integrating reinforcement learning models. Most importantly, CityFlow enables large-scale, interactive traffic simulations and opens new possibilities for advancing intelligent transportation systems through machine learning. In \cite{wei2019traffic} the CityFlow is used to develop the algorithm. The resources of this simulator can be found here: \url{https://github.com/cityflow-project/CityFlow}

\begin{figure}[h]
    \centering
    \includegraphics[width=0.5\linewidth]{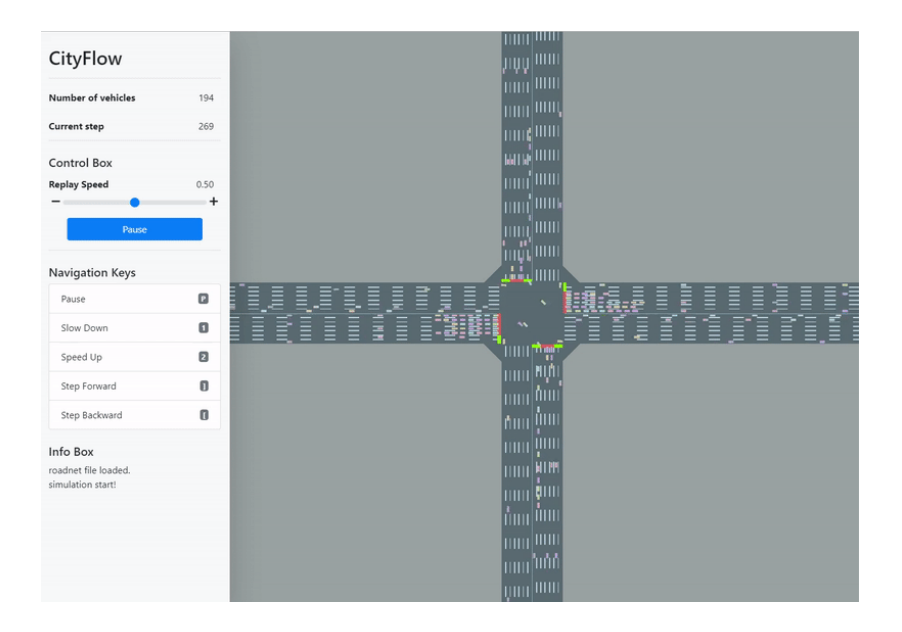}
    \caption{Snapshot from the CityFlow simulation which manages traffic flows allocation at the intersection for single agent RL and multiple intersection for MARL}
    \label{fig:enter-label}
\end{figure}

\subsection{Container Management}
\begin{figure}[h]
\centerline{\includegraphics[height=7cm]{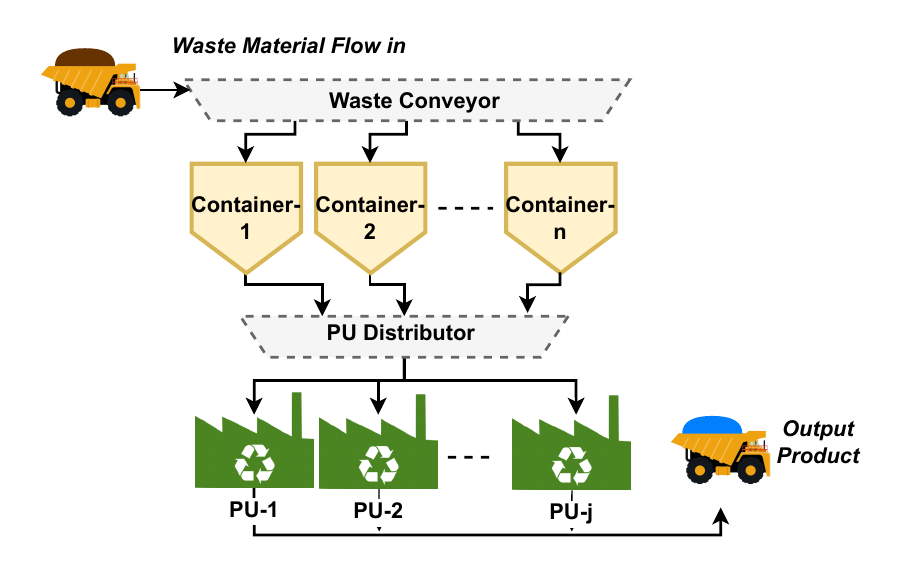}}
\caption{ContainerGym environment: This environment simulate the waste processing unit that is assumed the waste materials flow input continuously and store them to multiple containers. It should maintain the container level and allocate to Processing Unit (PU) that requiring waste materials to be processed.}
\label{fig:containerGym}
\end{figure}

A real-world industrial control task inspired resource allocation environment has been proposed in \cite{Pendyala2024contGym}. The authors outline the real-world industrial control task that served as the basis for RL benchmark. This task stems from the final phase of a waste sorting process (see Fig. \ref{fig:containerGym}). The environment comprises a solid material transformation facility containing multiple containers and a much smaller number of Processing Units (PUs). These containers are continuously filled with material, where the flow rate of the material follows a stochastic process that varies by container. Although this environment is developed for RL algorithm, it can be extended into the MARL settings by adding more agents. This work has a publicly available codes here: \url{https://github.com/Pendu/ContainerGym}.

\paragraph{Tested Algorithms}
Based on the available simulators, several algorithms have been found in the repository and reported in the literature as seen in Table \ref{tab:algorithmBenchmark}.
\begin{table}[h]
\caption{Summary of algorithms tested in the benchmarks or simulators}
\label{tab:algorithmBenchmark}
\begin{tabular}{ll}
\hline
\multicolumn{1}{c}{\textbf{Benchmark}} & \multicolumn{1}{c}{\textbf{Algorithm}} \\ \hline \hline
\textbf{BSK-RL}                        & PPO                                    \\
                                       & MAPPO + Communication               \\ \hline
\textbf{Power Distribution Networks}   & IAC                                    \\
                                       & IDDPG                                  \\
                                       & MADDPG                                 \\
                                       & SQDDPG                                 \\
                                       & IPPO                                   \\
                                       & MAPPO                                  \\
                                       & MAAC                                   \\
                                       & MATD3                                  \\
                                       & COMA                                   \\
                                       & FACMAC                              \\ \hline
\textbf{ContainerGym}                  & PPO                                    \\
                                       & TRPO                                   \\
                                       & DQN                                    \\ \hline
\textbf{Traffic Management}            & Graph based MARL                       \\ \hline
\end{tabular}
\end{table}

\section{Future Directions and Potential Challenges}
\subsection{Future Directions}
The application of MARL in RAO holds immense promise, with the potential to revolutionize numerous industries. As industrial systems become increasingly complex, decentralized, and dynamic, traditional resource allocation methods struggle to keep up. MARL’s capabilities, such as distributed decision-making, real-time adaptation, and modeling intricate agent interactions, position it as a key enabler for future resource management systems.

A primary area of future research is enhancing scalability and efficiency. With the growing scale of systems, such as the Internet of Things (IoT), edge computing, and smart cities, MARL must handle more agents and larger environments. Techniques like hierarchical MARL, mean-field approximations \cite{yang2018mean}, and decentralized learning frameworks \cite{zhang2019efficient} will be critical in addressing computational complexity and communication overhead in such large-scale systems.

Dynamic and autonomous systems will also benefit from MARL advancements. For instance, next-generation wireless networks (e.g., 6G) \cite{du2022multi}, energy-efficient resource management \cite{shen2022multi}, and autonomous transportation systems can leverage MARL to enable real-time and adaptive resource allocation. These applications demand decentralized learning capabilities and the ability to optimize system performance while ensuring sustainability.

Inter-agent coordination and communication remain vital research areas. As systems grow to be more complex, seamless collaboration between agents becomes increasingly important. Advances in coordination mechanisms, such as graph-based models \cite{zhang2023deepmag} and communication-free methods, will improve resource sharing in cooperative environments while ensuring fairness in competitive settings.

Additionally, real-world resource allocation often involves non-stationary environments and heterogeneous agents \cite{zhong2024heterogeneous}. Future MARL research will focus on enhancing agent adaptability to evolving conditions and diverse objectives. By improving generalization across environments, MARL systems can rapidly adapt to new scenarios, making them more practical for dynamic resource allocation challenges.

Finally, integrating MARL with emerging technologies like federated learning \cite{zhang2021optimizing,li2022federated} and quantum computing \cite{yun2023quantum} presents exciting opportunities. These integrations could enable secure, efficient, and decentralized resource management systems, transforming domains like energy, logistics, and beyond by ensuring both efficiency and resilience in the face of growing complexity.

\subsection{Potential Future Challenges}
MARL offers a powerful solution for addressing the complexity and dynamic nature of RAO. By allowing decentralized agents to learn optimal strategies through interaction with their environment, MARL can handle the challenges of non-stationary condition, competition, and cooperation, providing a flexible, scalable approach for optimizing resource distribution in a variety of domains. However, several challenges remains in this domain such as: 
(1) Partial Observability \cite{Baker2020Emergent}: Agents often have limited information, requiring approaches that can handle partial observability.
(2) Convergence \cite{yu2022surprising}: Careful algorithm design and hyperparameter tuning are needed to ensure convergence in competitive scenarios.
(3) Scalability \cite{yang2018mean,wang2020large}: Techniques like mean field approximations are used to handle large numbers of agents.
(4) Safety Constraints \cite{lu2021decentralized}: Implementing safe exploration techniques to avoid harmful resource allocations during learning.
(5) Communication Overhead \cite{10.5555/3635637.3663308}: Centralized approaches require a high level of communication between agents and the central controller. This can lead to bottlenecks, especially in real-time environments or large-scale systems.
(6) Exploration \cite{cui2019multi}: The joint exploration of the state-action space by all agents is more complex than in decentralized methods.

\section{Conclusion}
This survey examines the intersection of MARL and RAO, highlighting recent advances and emerging trends in this rapidly evolving area. MARL has demonstrated considerable potential in addressing the challenges of decentralized and dynamic resource allocation by enabling agents to learn and adapt in complex, uncertain environments. We reviewed core methodologies, surveyed applications across diverse domains, and analyzed the strengths and limitations of existing approaches.

Despite this progress, several key challenges remain, including non-stationary, limited scalability, coordination complexity, and the need for more generalizable algorithms. Tackling these issues calls for further research into improved training paradigms, adaptive communication mechanisms, and hybrid techniques that integrate MARL with classical optimization strategies.

Future work should also prioritize the deployment of MARL in emerging domains, the establishment of standardized RAO benchmarks, and the development of task-relevant evaluation metrics. By addressing these gaps, the community can advance the capabilities of MARL and broaden its impact on real-world resource allocation systems.

\backmatter
\bmhead{Acknowledgments}
This work has been supported by the SmartSat CRC, whose activities are funded by the Australian Government’s CRC Program.

\bibliographystyle{bst/sn-chicago}
\bibliography{references}

\end{document}